\documentclass[reprint,amssymb, amsmath, nobibnotes, nofootinbib, aps, superscriptaddress]{revtex4-1}

\usepackage{graphicx}

\usepackage[unicode=true, pdfusetitle,
 bookmarks=true,bookmarksnumbered=false,bookmarksopen=false,
 breaklinks=false,pdfborder={0 0 0},backref=false,colorlinks=false]
 {hyperref}

\begin{document}

\title{Probing Ultrafast Dynamics with Time-resolved Multi-dimensional\\Coincidence Imaging: Butadiene}

\author{Paul Hockett}
\email{paul.hockett@nrc.ca}
\affiliation{National Research Council of Canada, 100 Sussex Drive, Ottawa, Ontario, Canada}
\author{Enrico Ripani}
\affiliation{Dipartimento di Chimica, Università La Sapienza, Piazzale Aldo Moro 5, Rome, Italy}
\author{Andrew Rytwinski}
\affiliation{Department of Chemistry, Queen's University, Kingston, Ontario, Canada}
\author{Albert Stolow}
\affiliation{National Research Council of Canada, 100 Sussex Drive, Ottawa, Ontario, Canada}
\affiliation{Department of Chemistry, Queen's University, Kingston, Ontario, Canada}

\date{\today}

%

\begin{abstract}

\begin{description}

\item[Status]\textit{To appear in JMO special issue ``Ultrafast Dynamical Imaging of Matter", summer 2013.}
\item[Version]\textit{RevTex~4.1 typeset version for arxiv.}

\end{description}

Time-resolved coincidence imaging of photoelectrons and photoions
represents the most complete experimental measurement of ultrafast
excited state dynamics, a multi-dimensional measurement for a multi-dimensional
problem. Here we present the experimental data from recent coincidence
imaging experiments, undertaken with the aim of gaining insight into
the complex ultrafast excited-state dynamics of 1,3-butadiene initiated
by absorption of 200~nm light. We discuss photoion and photoelectron
mappings of increasing dimensionality, and focus particularly on the
time-resolved photoelectron angular distributions (TRPADs), expected
to be a sensitive probe of the electronic evolution of the excited
state and to provide significant information beyond the time-resolved
photoelectron spectrum (TRPES). Complex temporal behaviour is observed
in the TRPADs, revealing their sensitivity to the dynamics while also
emphasising the difficulty of interpretation of these complex observables.
From the experimental data some details of the wavepacket dynamics
are discerned relatively directly, and we make some tentative comparisons
with existing \emph{ab initio} calculations in order to gain deeper
insight into the experimental measurements; finally, we sketch out
some considerations for taking this comparison further in order to
bridge the gap between experiment and theory.
%
\end{abstract}

\maketitle

\section{Introduction\label{sec:Introduction}}

Coincidence imaging techniques, in which the full momentum vector
of both photoelectron and photoion is measured, have been growing
in popularity and sophistication over the last 20 years. The COLTRIMS
(Cold Target Recoil Ion Momentum Spectroscopy) community, in particular,
has developed significant expertise in relatively high-energy and
multi-coincidence measurements \cite{Moshammer1994,Moshammer1996,Ullrich1997,Dorner2000,Czasch2005},
typically (although not exclusively) utilising synchrotron light sources
in photoionization studies. The original aim of COLTRIMS was application
to many-body collision dynamics, via kinematically complete measurements
of collision systems \cite{Dorner2000}, although the technique has
since been applied to studies as diverse as photoelectron diffraction
\cite{Landers2001}, probing entanglement \cite{Akoury2007} and the
investigation of tunnel ionization \cite{Meckel2008,Eckle2008}, as
well as extensive studies of photoelectron angular distributions \cite{Ueda2003,Reid2012}.
Flat-field and VMI based coincidence imaging experiments have also
been used by a handful of groups \cite{Davies1999,Lafosse2000,Continetti2001,Lebech2002,Gessner2006,Vredenborg2008,Bisgaard2009},
usually with a focus on lower-energy processes as appropriate to experiments
based around table-top laser sources. The first demonstration of femtosecond
time-resolved coincidence imaging to study photochemical processes
was over a decade ago \cite{Davies1999,Davies2000}, and this type
of imaging measurement provides the fullest experimental dataset possible,
which can be considered as a 7D measurement, or even 8D if one considers
the fragment mass spectrum as a distinct observable to the fragment
velocity distributions; time-resolved coincidence imaging therefore
provides the best chance of elucidating complicated, multi-dimensional,
excited state dynamics from experimental measurements. 

Despite the potential of coincidence imaging, the technique has had
only a small impact thus far to time-resolved measurements generally,
and more specifically to measurements utilising UV sources \cite{Reid2012}.
The difficulty of applying coincidence imaging in time-resolved, UV
pump-probe experiments is partly due to the limitation of single-particle
counting techniques - with consequent requirements for long experimental
runs and long-term experimental stability - which makes time-resolved
experiments particularly challenging; additionally there is the inherent
difficulty of producing and controlling short-pulse UV light. A particular
issue is the minimisation of background signal from scattered light,
which becomes a problem on a per photon basis once photon energies
are above the work function of the materials used in the spectrometer
\cite{Lee2007,Liu2011,clarkin2012,owenScattLight2012}. The benefit
of UV wavelengths is that, for many small molecules, the photon energies
are sufficient for 1-photon pump, 1-photon probe experimental schemes,
which are ideal for the study of the dynamics of electronically excited
states of molecules. In these kind of schemes the laser intensities
can be kept low and well within the perturbative regime ($\ll10^{12}$~Wcm$^{-2}$),
and the observables take their simplest form. The few successful studies
to date \cite{Davies1999,Rijs2004,Gessner2006,Bisgaard2009} have
begun to explore the power of the technique, but much work remains
to be done.

The observables provided in a full imaging study provide additional
information beyond the energy-time mapping of a, now routine, 2D photoelectron
or photoion measurement, which provide time-resolved photoelectron
spectra (TRPES) or mass spectra (TRMS) respectively. In particular
the time-resolved photoelectron angular distributions (TRPADs) are
sensitive to the electronic structure of the ionizing state, so are
expected to reveal subtle details of the non-adiabatic electronic
dynamics \cite{Arasaki2000,Seideman2002,Suzuki2003,Stolow2008,Arasaki2010}.
In the simplest case one might concoct, that of passage through a
conical intersection (CI) leading to a change in the electronic symmetry
of the excited state, the PAD is expected to change reflecting the
non-adiabatic dynamics and map principally the electronic part of
the dynamics \cite{Suzuki2003,Wu2011}. This expectation can be contrasted
with the time-resolved photoelectron spectrum, which can be considered
to be an observable dominated by vibrational motions \cite{Wu2011},
but may additionally map some aspects of the non-adiabatic electronic
dynamics cleanly depending on the ionization correlations \cite{Blanchet1999,Blanchet2001,Schmitt2001}.

However, in the case of large amplitude motions on a single electronic
state (adiabatic dynamics) changes in the PAD are also expected and,
furthermore, the PAD is energy dependent, so changes to the vertical
ionization potential (IP) as a function of nuclear coordinates will
also couple into the form of the observed PAD. These factors make
the mapping of the dynamics onto the TRPADs non-trivial to understand
at both a qualitative and quantitative level, and obviate the (relatively)
simple picture that the TRPADs map only the electronic dynamics for
all but the simplest of cases; recent computational studies of excited
state dynamics in NO$_{2}$ \cite{Arasaki2010} have illustrated the
response of the TRPES and TRPADs to complicated excited-state dynamics,
demonstrating the richness of the observable while also suggesting
the difficulty of obtaining detailed insights into molecular dynamics
via a purely experimental approach, with no \emph{a priori} knowledge
of the underlying dynamics. Of particular note is the non-isomorphic
nature of the nuclear configuration and observable mapping spaces:
the wavepacket motion on the excited-state, which includes dispersion,
bifurcation, interferences and other complex quantum-mechanical behaviours,
does not allow for a direct mapping of a given dimension in nuclear
coordinate space onto a given dimension of the observable (e.g. photoelectron
energy, anisotropy parameter), although such mappings may be possible
in low dimensionality problems such as vibrational wavepackets in
diatomics \cite{Shapiro1999,Arasaki2000,Arasaki2000a,Wollenhaupt2005}.
However unsurprising this conclusion is, the knowledge gap between
the observable and the underlying wavepacket dynamics is often overlooked
or ignored in treatments of time-resolved measurements.

Despite these difficulties, the TRPADs provide an additional observable
- therefore more information - than the TRPES alone, and additional
dimensionality to the dataset. One demonstration of the utility of
this higher information content has been the interpretation of experimental
measurements via qualitative/semi-quantitative modelling of TRPADs,
which can provide insight into the mapping of the excited state wavepacket
to the observables without the need for full \emph{ab initio} treatments,
and yield deeper insight into the molecular dynamics than the TRPES
alone \cite{Gessner2006,Hockett2011}. Many studies based on velocity
map imaging (VMI) measurements have also illustrated the utility of
TRPADs at a phenomenological level (in no small part because PADs
come {}``for free'' with the technique) and a recent review article
has surveyed much of the work by such {}``users'' of photoelectron
angular distributions \cite{Reid2012}. 

The ultrafast dynamics of 1,3-butadiene ($C_{4}H_{6}$) have been
studied in considerable detail experimentally and theoretically, most
recently by Boguslavskiy et. al. \cite{bogusNatChem} and Levine et.
al. \cite{Levine2009} (for a more detailed overview of the butadiene
literature to date see ref. \cite{Levine2009}). In the experimental
photoelectron study, butadiene was excited to the bright $1{}^{1}B_{u}$
state, with UV radiation around 216~nm (5.74~eV), and probed via
ionization with a time-delayed UV pulse at 266~nm (4.66~eV). The
computational studies, based around an \emph{ab initio} multiple spawning
(AIMS) methodology, focussed on describing the excited state dynamics
of population on the same bright state and with similar energy to
the experimental case.%
\footnote{In this case neither the experimental pump bandwidth, nor the probe
pulse, were included in the simulations. More recently the AIMS methodology
has been expanded to explicitly include the probe process, although
this has so far only been applied to the photoionization of ethylene
with VUV pulses \cite{Tao2011}.%
} The conclusions from these complementary experimental and theoretical
studies are in accord, and point to rapid and complicated dynamics
involving fast motion on the initially populated bright $1{}^{1}B_{u}$
state (historically termed the $S_{2}$ state, as it lies higher in
energy in the Franck-Condon region), with twisting about the carbon
backbone and out-of-plane bending motions leading to highly distorted
geometries (relative to the planar ground state) on $<$40~fs timescales
\cite{Levine2009}. At least two minimum energy conical intersections
(CIs), coupling $S_{2}$ to $S_{1}$ (the optically dark $2^{1}A_{g}$
state) and three CIs coupling $S_{1}$ to $S_{0}$, were found to
play important roles in the relaxation of the excited state. In a
wavepacket picture, the initial dynamics from the Franck-Condon region
to the first CI would correspond to rapid passage down steep gradients,
with little dispersion of the wavepacket along other coordinates.
Once on $S_{1}$, the topology would cause the wavepacket to split,
with parts heading towards each CI, and there would be the possibility
of more complex dynamical behaviour. Away from the CIs, the non-adiabatic
coupling of the $S_{2}$ and $S_{1}$ states is strong over large
regions of the nuclear configuration hyperspace, with significant
interaction between the states even at the equilibrium geometry -
for example, the {}``dark'' $2^{1}A_{g}$ state is actually found to carry
non-negligible oscillator strength due to non-adiabatic coupling with
the bright state \cite{Levine2009}. The strongly coupled nature of
these states and short timescales involved suggests that butadiene
can be considered as something of a prototypical, perhaps even limiting
or pathological, case for rapid dynamics; as such butadiene is a good
exemplar of a system where the information available from frequency
resolved measurements is very limited,%
\footnote{For instance ref. \cite{Leopold1984} discusses the experimental absorption
spectra, and the lack of information obtainable from the very diffuse
bands observed (with just three broad features visible over the 44000~-~51000~cm$^{-1}$
region studied); ref. \cite{Krawczyk2000} provides details of \emph{ab
initio} calculations of the optical spectra which require the inclusion
of a phenomenological dephasing constant to match the experimental
data, again indicating a gap in the understanding of the radiationless
relaxation of the excited state.%
} but the speed and complexity of the dynamics make time-resolved measurements
technically demanding and difficult to interpret.

Experimental measurements have probed the projection of these dynamics
onto the time-resolved ion yields \cite{Assenmacher2001,Fuss2001}
and time-resolved photoelectron spectrum \cite{bogusNatChem}. In
the former case fast time-constants, $<$50~fs, were determined for
the decay of the parent ion signal, and delayed onset of fragmentation
was also observed \cite{Assenmacher2001}; in the latter case, as
well as a rapid decay of the photoelectron yield, a fast shift in
the vertical ionization potential and a photoelectron band with little
structure were observed. Recent work, in which the TRPES was measured
in coincidence with the mass spectrum \cite{bogusNatChem}, has demonstrated
the separation of the photoelectron spectra correlated with the $S_{2}$
and $S_{1}$ states due to the ionization correlations with $D_{0}$
and $D_{1}$ ion states respectively; the $D_{0}$ state is stable,
yielding parent ion signal only, while the $D_{1}$ state fragments
\cite{Werner1975,Dannacher1980,Fang2011}, thus producing a very different
mass spectrum. In this way coincidence measurements have been able
to help disentangle the TRPES data by providing complementary, correlated
observables. The same photofragment coincidence technique has also
been applied to strong-field ionization of butadiene in order to probe
multi-electron effects \cite{Boguslavskiy2012}. Measurement of the
TRPES and time-resolved mass spectrum (TRMS) in coincidence provides
a 3D dataset, with the additional dimension of ion time-of-flight
and, depending on the details of the measurement, may also provide
information on the kinetic energy release of the fragments. One might
hope, therefore, that a full, 7D, coincidence imaging experiment provides
a rich enough dataset to discern some specific, possibly mechanistic,
details of excited-state dynamics in molecular systems, even in the
case of very fast dynamics exemplified by butadiene.

In order to explore some of these issues we present here our recent
experimental results, focussing on a presentation of a full dataset
to illustrate the richness of the 7D coincidence measurements. Along
with the data, we include a brief description of our coincidence imaging
apparatus, and a qualitative analysis of the data, with a focus on
the TRPADs. In future publications, aspects of our apparatus will be
discussed in fuller detail, and a more detailed interpretation of
the experimental data and comparison with recent ab initio dynamics
calculations will be made.

\section{Experimental}

Time-resolved pump-probe measurements were carried out with sub-40~fs
UV pulses, $\lambda_{pump}=$200~nm (6.20~eV) and $\lambda_{probe}=$266~nm
(4.66~eV). Butadiene (1\% in He) was introduced to the interaction
region via a 1~kHz pulsed valve (Even-Lavie \cite{Even2000}, 150 $\mu$m
diameter conical nozzle) with a stagnation pressure of $\sim$5~bar.
Full 7D coincidence measurements were performed on a coincidence imaging
spectrometer (CIS). Details of the apparatus are provided in the following
sections.

\subsection{Optical set-up\label{sub:Optical-set-up}}

\begin{figure}
\begin{centering}
\includegraphics{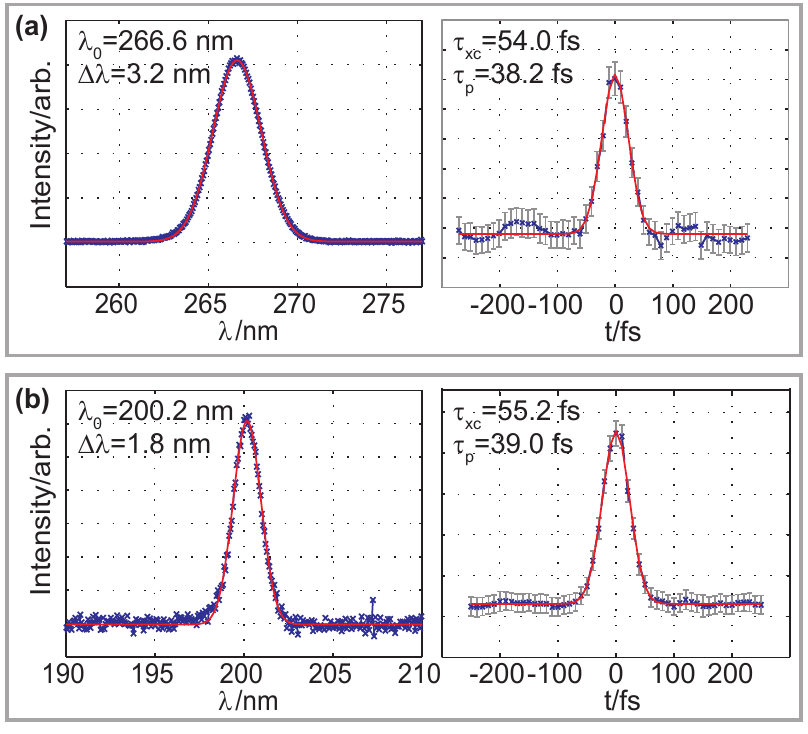}
\par\end{centering}

\caption{Spectra and autocorrelation traces for the (a) 266~nm probe pulses,
(b) 200~nm pump pulses. Spectra show central wavelength and FWHM
for the Gaussian fit. Autocorrelation traces show $\tau_{XC}$ (the
FWHM of the Gaussian fit), and the corresponding pulse $\tau_{p}$
duration assuming a Gaussian envelope.\label{fig:spectra}}

\end{figure}

Short pulse infrared light ($\lambda$=800~nm, 35~fs, 1~kHz repetition
rate) was generated by a standard titanium-sapphire based regenerative
amplifier system, followed by a single pass amplifier stage (Coherent Legend
Elite Duo). Approximately 700~$\mu J$ of the output was used to
pump a 3$^{rd}$ ($3\omega$) and 4$^{th}$ ($4\omega$) harmonic
generation scheme, based on sum-frequency mixing in thin BBO crystals.

Calcium fluoride prism pairs were used to compress the generated UV
pulses, and compensate for the dispersion of transmissive optics in
the beam paths (harmonic separation and recombination optics, $\lambda/2$
plate for 3$^{rd}$ harmonic, experimental chamber window, propagation
in air). The output pulses were measured directly with an autocorrelator
based on 2-photon absorption \cite{Homann2011}, and the pulse durations
in the experimental chamber were additionally confirmed via the cross-correlation
feature obtained from fitting the TRPES data. Output pulses on the
order of 35~$\pm$~3~fs were measured. Typical autocorrelation
traces, along with pulse spectra measured with a UV spectrometer (Ocean
Optics Maya Pro) are shown in figure \ref{fig:spectra}.  

Control over pump-probe delay was achieved via a high-precision linear
motor stage (Newport XML210), for the data reported herein steps of
10~fs were used. The beams were recombined in a collinear geometry,
sent through a spatial filter to clean the mode and increase the beam
diameters by a factor of 2, and loosely focussed into the interaction region
via an $f$=~1~m focussing mirror. Beam diameters of the UV at the
focussing mirror were $~$4~mm at $3\omega$ and $~$2.6~mm at $4\omega$.
Focal spot sizes in the interaction region were estimated to be $\sim$85~$\mu$m
at $3\omega$ and $\sim$100~$\mu$m at $4\omega$. To avoid multi-photon
effects, and minimise scattered light signal from the $4\omega$ light,
relatively low pulse energies of $\sim200\, nJ$ at $3\omega$ and
$\sim10\, nJ$ at $4\omega$ were used, corresponding to peak intensities
of $\sim$1.9x10$^{11}$~Wcm$^{-2}$ and $\sim$7x10$^{9}$~Wcm$^{-2}$
respectively.

\subsection{Coincidence imaging spectrometer\label{sub:Coincidence-imaging-spectrometer}}

\begin{figure}
\begin{centering}
\includegraphics{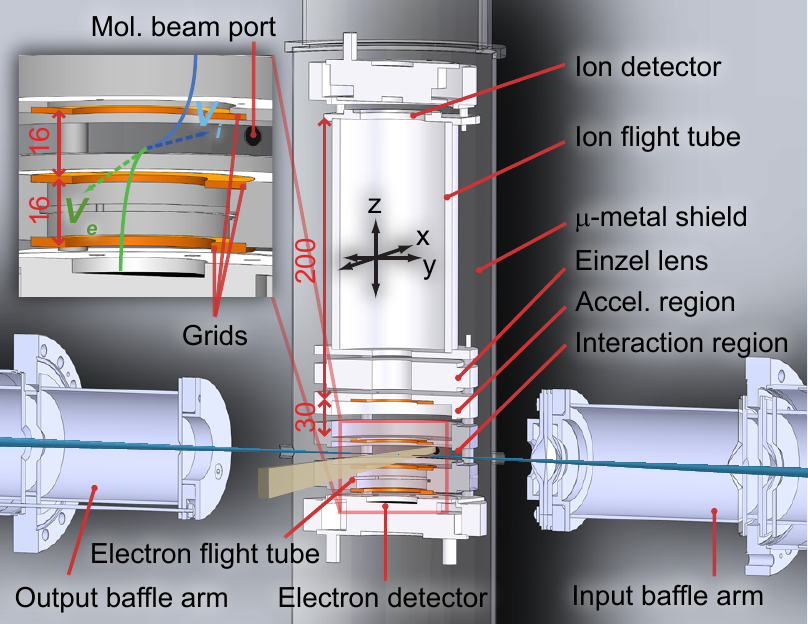}
\par\end{centering}

\caption{Schematic of the coincidence imaging spectrometer, showing key aspects
of the design. The inset shows details of the interaction region and
electron flight region, including electron and ion trajectories; dashed
lines represent the initial velocity vectors and the solid lines the
paths in the guiding fields. Dimensions shown are in mm, and the axis
definitions shown are used throughout this work.\label{fig:CIS-schematic} }

\end{figure}

The coincidence imaging spectrometer (CIS) is illustrated in figure
\ref{fig:CIS-schematic}. An overview is provided in the following,
and a more detailed description may be found in ref. \cite{Lee2007}.
The imaging of both electrons and ions is based on a flat-field, Wiley-McLaren
type geometry. A flat-field design was chosen to allow for a large
turn-around time spread of photoelectrons along the time-of-flight
($ToF$) axis (labelled as the $z$-axis in figure \ref{fig:CIS-schematic}),
as compared with a VMI type configuration, and also to provide the
most direct mapping of $(x,y,ToF)$ data to initial velocity vector,
thus allowing for a simple calibration and data backtransformation
procedure. To create flat fields, the interaction region is enclosed
by grids in the vertical direction; the open apertures in the horizontal
plane, required to admit the supersonic molecular beam and laser beams,
are removed as far from the interaction centre as possible and minimised
in spatial extent in order to avoid aberrations and fringing fields.
The bottom grid is biased (typically +10~to~+20~V) relative to
the top grid (0~V) in order to extract photoelectrons towards the
lower flight tube. After the photoelectrons have cleared the interaction
region a high voltage pulse (on the order of -0.5 to -0.8~kV, $<$15~ns
rise time, supplied by a HVC-1000 pulser unit from GPTA) is applied
to the top grid to eject the photoions to the upper detector. The
photoion flight tube also contains an acceleration region, and an
Einzel lens assembly (not used in this work), for further control
over the ion imaging spectrometer conditions. The pulsed valve (not
illustrated) is situated in a source chamber, separated from the CIS
chamber by a conical molecular beam skimmer (Beam Dynamics, 1~mm
aperture), approximately 45~cm from the interaction
region. The molecular beam is further skimmed by an adjustable slit, discussed
below. Partly shown in figure \ref{fig:CIS-schematic} are the baffle
arms, designed to limit the background photoelectron signal from scattered light,
a particularly severe problem from the 200~nm pump pulse. Further
details of the scattered light problem and the baffle system employed
here will be given in a future publication \cite{owenScattLight2012}.

Both detectors are comprised of a triple stack of 40~mm diameter
MCPs and delay line anodes (Sensor Sciences LLC). Time-of-flight measurement
is made via a pick-off from the MCP front face, and $(x,y)$ position
data is obtained from the delay line signals. The ion detector is
offset slightly (15~mm) along the molecular beam axis to compensate for
the initial translational velocity along this axis. The data read-out
and storage is handled by an electronics rack comprising NIM modules
on a CAMAC bus backbone, for full specifications see ref. \cite{Lee2007},
with final output of the processed signals to a PC. The timing-resolution,
averaged over the full detector area, is $\sim$200~ps, limited by
the pulse propagation characteristics over the detector face; the spatial resolution
is $<$80~$\mu$m for both $x$ and $y$ axes of the electron detector,
and $<$60~$\mu$m for the ion detector \cite{Lee2007}. The PC also
controlled other experimental variables such as the pump-probe delay,
the shutters installed in the pump and probe beam paths, and dwell
times at each delay. Data was measured with dwell times defined by
events (as opposed to laser shots), ensuring good Poissonian statistics at all
pump-probe time delays even when absolute count rates were low. Pump-only and probe-only
signals were measured at each delay for dwell times of $1/N$ less
than the pump-probe signal, where $N$ was the total number of delays
set. Each set of $N$ delays defines one experimental cycle, and the
final dataset presented here comprises $\sim200$ cycles; cycles were
kept short ($\lesssim$30 minutes of real-time) to minimise the effects
of any drifts during the course of a cycle.

Instrument resolution is determined both by the overall mechanical
design, optimised for low-energy photoelectrons and photoions ($\lesssim$2~eV),
the acquisition electronics and the fields applied for a given measurement
\cite{Lee2007}. Adjustment of the extraction fields provides experimental
control over $(x,y)$ resolution via the choice of the energy cut-off
(image size), in essence higher extraction fields increase the dynamic
range of the image at the cost of energy resolution, while lower extraction
fields maximise energy resolution over a reduced dynamic range. This
consideration is common to VMI and other imaging techniques \cite{Parker1997,sevi2004,powis2005,hockettThesis2009}.
In 3D imaging, in common with 1D time-of-flight measurements, the
temporal resolution ($z$-axis) is also affected by the choice
of extraction fields due to the influence of the fields on the turn-around
time-spread of the particles, hence temporal spread at the detector
\cite{Chichinin2009}. Under typical operating conditions, with photoelectrons
up to 1~eV extracted with a +13.5~V field, the photoelectron energy
resolution $\Delta E/E$ was calculated from SIMION simulations to
be 1\% (10~meV) for the $x$-axis, 10\% (100~meV) for the $y$-axis
and 3\% (30~meV) for the ToF axis \cite{Lee2007}. Similar figures
of 1\%, 15\% and $<$1\% were calculated for the ion $(x,y,ToF)$
resolution at 0.78~eV with a 200~V extraction field and 600~V acceleration
field \cite{Lee2007}. In practice the $y$-resolution was improved
from these design stage simulations by a reduction of the ionization
volume along the laser axis, reducing $\Delta y$ from the 2~mm assumed
in the resolution figures provided above. This reduction was achieved
by creating a pseudo-1D molecular beam source along the $y$-axis,
via the use of a piezo-actuated razor blade slit, mounted after the
skimmer which separates the source and spectrometer chambers. In the
experiments detailed herein the slit width was set at 400~$\mu$m, providing
a significant improvement on the $y$-axis resolution and allowing
for energy slices of 100~meV to be used throughout the photoelectron
data analysis over the full 2~eV energy range imaged.

\subsection{Data analysis \& calibration\label{sub:Data-analysis-and-calib}}

For each event - photoelectron and/or photoion detection - a data
record is stored, consisting primarily of position and time-of-flight
$(x,y,ToF)$, and anode charges $(Q_{x},Q_{y})$, for the ion and
electron. Also stored for each data record are various indices allowing
correlation of the event with pump-probe configuration, pump-probe
delay, experimental cycle etc. In the current electronics configuration
only a single electron and/or ion event is recorded per laser shot,
although in principle multi-hit operation of the ion detector is possible
and only limited by the dead-time of the detector, as defined by the
time taken for pulses to clear the delay lines ($\lesssim$15~ns).
The maximum count rate for electrons is therefore the limiting factor,
and is the same as the repetition rate of the laser (1~kHz), although
for operation in a true coincidence regime lower count rates may be
required to limit false coincidence events to a reasonable level \cite{Stert1999,Lee2007,jochenStats2012}.
In the experiments presented here the count rates were kept to $\lesssim$100~Hz
for the dominant parent ion channel, which should limit false coincidences
to $<$10\% \cite{Lee2007}. It is interesting, although unsurprising,
to note that higher total count rates are permissible in a channel-resolved
experiment as compared to a single channel case \cite{Stert1999}.
For example, the majority of background electrons from scattered light
appear at early $ToF$ relative to the main signal electrons, so are
temporally resolved and do not contribute to false coincidences. A
more thorough statistical analysis of the multi-channel case will
be presented in another publication \cite{jochenStats2012}.

The data hypercube obtained experimentally may be filtered along any
or all dimensions in order to examine correlations, throw out bad
data, retrieve various mappings of the data and so on. For example,
charge histograms allow for the determination of a window of good
events, defined as single hit events, and the rejection of bad events
where two or more hits were registered within the delay line pulse
transit period, such bad events have higher $(Q_{x},Q_{y})$ than
single hit events. Correlation of photoelectrons with photoions of
high translational velocity eliminates signal from ionization of background
gas (see section \ref{sub:Expt-Overview}); correlation with a given
mass provides fragment-resolved data, and so forth.

Calibration and backtransformation of the data is performed via a
three-step process: (1) the image centre $(x_{0},y_{0},z_{0})$ is
defined, either by manual inspection, or taken as the peak in the
$(x,y,ToF)$ histograms; (2) the $(x,y)$ position bins are converted
to mm positions from the image centre, based on a (static) look-up
table calibration, and converted to velocities $(V_{x},V_{y})$ by making use
of the measured time-of-flight; (3) a $ToF$ look-up table, based
on numerical trajectory calculations which account for the voltages
applied, is computed and applied to convert the $ToF$ data to $V_{z}$.
Because of the flat-field arrangement the trajectory calculations
are straightforward and not computationally demanding.

Once converted to velocity space, energy and angular data can readily
be extracted from the dataset. Obtaining these integrated observables
is again just a case of creating histograms of the various quantities
of interest, combined with filtering as described above, to provide
maps of observables for given regions of the data hypercube. The histogram
of events with energy ($E=\frac{1}{2}m\mathbf{V}^{2}$) versus time
delay ($t$) provides the 2D map ($E,\, t$) - the TRPES - with bins
of width $\Delta E$ and $\Delta t$. Conversion of the data to spherical
polar coordinates $(E,\theta,\phi)$ allows binning into volume elements
$\Delta E\sin\theta\Delta\theta\Delta\phi$ for each delay $t$,
to create 3D maps of the energy and angular distributions of events.
Although the full, quasi-continuous, 3D distribution may be visualised
as a set of isosurfaces or projection planes from these maps, the
data is perhaps represented most tractably as a series of TRPADs at
selected energy and time slices, with angular distribution $I(\theta,\phi;\, E,\, t)$.
For cylindrically symmetric distributions, further integration over
$\phi$ can also be performed leading to a reduced form $I(\theta;\, E,\, t)$.
Low event number datasets also benefit from this integration because
without integration or smoothing the 3D maps may be too sparse or
noisy for further analysis.

The extracted angular distributions can also be described phenomenologically
by $\beta_{LM}$ parameters, 

\begin{equation}
I(\theta,\phi;\, E,\, t)=\sum_{L,M}\beta_{LM}Y_{LM}(\theta,\,\phi)\label{eq:IBlm}\end{equation}
where $Y_{LM}(\theta,\phi$) are spherical harmonics of rank $L$
and order $M$. This distribution is general to any scattering system
\cite{yang1948} but the expansion is constrained, by the symmetry
of the experiment, to only certain values of $L,\, M$ \cite{Reid2003};
in the case of the pump-probe experiment considered here, with total
absorption of two photons, linearly polarised light and pump and probe
polarisation vectors parallel, the final angular distributions contain
terms $L=0,\,2,\,4$ and are cylindrically symmetric ($M$=0) with
symmetry axis defined by the laser polarisation. For the photoelectrons,
the $\beta_{LM}$ contain details of the ionization dynamics, and
are complex functions of molecular geometry and photoelectron energy
\cite{Seideman2002,Reid2003,Stolow2008}. For the photoions, the $\beta_{LM}$
contain details of the fragmentation dynamics (although the expansion
is not usually written in this exact form, see for example refs. \cite{Dixon1986,Rakitzis1999,Clark2006,Suits2008,Rakitzis2010}).

\section{Results \& discussion}

In this section we present a complete dataset of photoelectron and
photoion measurements. An overview of the data is first given, providing
time-integrated 3D visualizations of the full datasets. Low dimensionality
(TRPES, TRMS) and high dimensionality (TRPADs, correlated data) maps
extracted from the data are then presented, followed by a brief discussion
of the results in the context of gaining insight into the underlying
molecular dynamics.

\subsection{Overview - visualizing multi-dimensional measurements\label{sub:Expt-Overview}}

\begin{figure}
\begin{centering}
\includegraphics{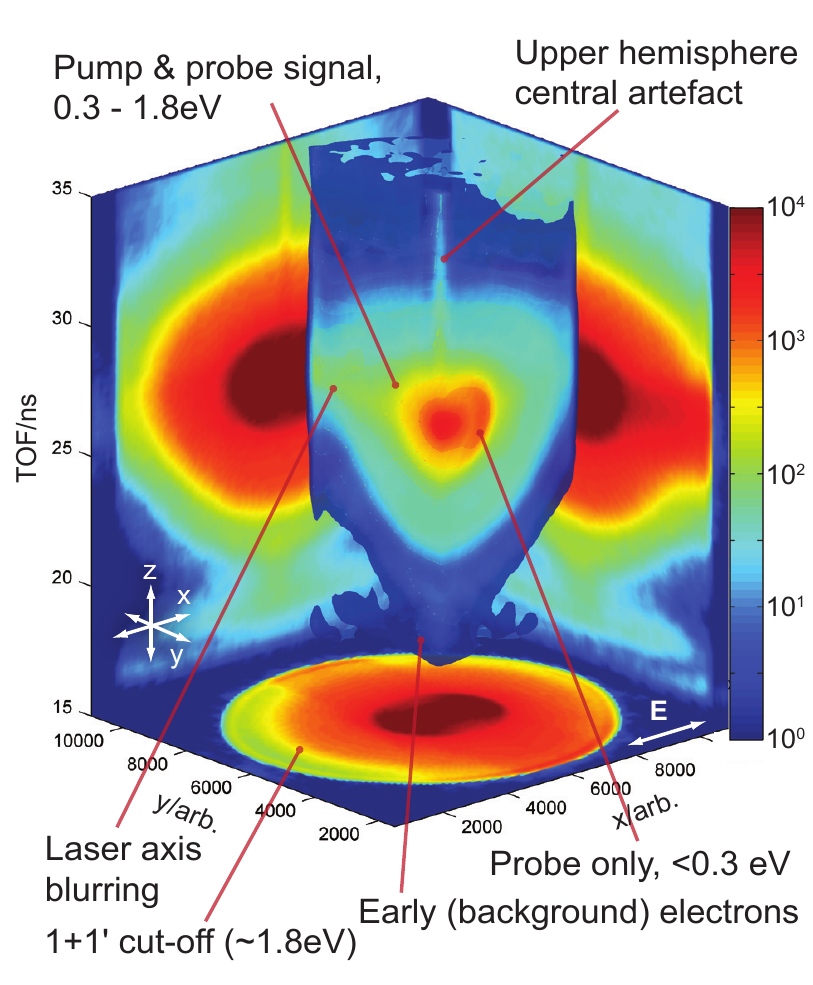}
\par\end{centering}

\caption{Electron imaging summary. One quadrant of the raw, time-integrated
3D electron data is shown as nested isosurfaces. The 2D image planes
show the $(x,y)$, $(y,z)$ and $(x,z)$ projections of the raw data. For the position data, 1~bin~$\approx$~5~$\mu$m. 
This is an interactive figure in some versions of this manuscript;
the interactive version, and source data, is also available at \protect\href{http://dx.doi.org/10.6084/m9.figshare.106343}{http://dx.doi.org/10.6084/m9.figshare.106343}.\label{fig:Electron-imaging-summary} }
\end{figure}

\begin{figure}
\begin{centering}
\includegraphics{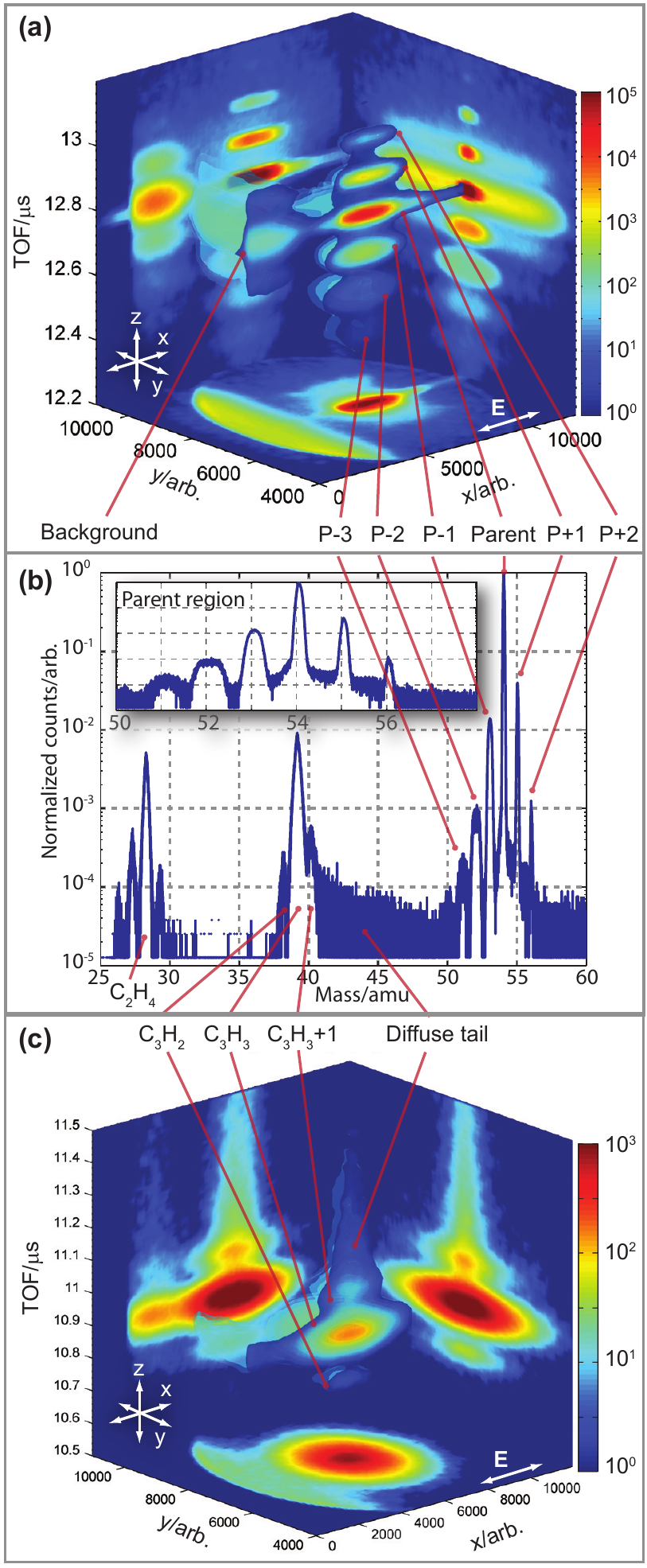}
\par\end{centering}

\caption{Ion imaging summary (time-integrated). (a) Raw 3D data $(x,y,ToF)$
for the parent ion region. Raw 3D data, sliced along the centre of
the distribution, is shown by the nested isosurfaces, and 2D projections
onto the $(x,y)$, $(y,z)$ and $(x,z)$ image planes are also shown.  For the position data, 1~bin~$\thickapprox$5~$\mu$m. 
Colour mapping shows log$_{10}$(counts), with a dynamic range of
$~$5 orders of magnitude. (b) Time-integrated mass spectrum. Obtained
by integrating the data over $(x,y)$ (background gas signal excluded).
The inset shows a detailed view of the parent ion region, 50~-~58~a.m.u.
(c) As (a) but for fragment region centred at 39 a.m.u. This is an
interactive figure in some versions of this manuscript; the interactive
version, and source data, is also available at \protect\href{http://dx.doi.org/10.6084/m9.figshare.106343}{http://dx.doi.org/10.6084/m9.figshare.106343}.\label{fig:Ion-imaging-summary}
}

\end{figure}

The most direct way to begin considering the data is via 3D maps of
the electron and ion signals. Here we present only the time-integrated
data in this form, although visualization of time-sliced data as 3D
maps is also a useful technique for qualitative analysis. The 3D maps
provide direct information on the performance of the instrument, a
quick check for artefacts, some information on the shape of the data
in terms of complexity, ion fragment yields, energy spectra and so
on; they represent a framework within which to begin a more detailed
analysis of the data.%
\footnote{Figures \ref{fig:Electron-imaging-summary} \& \ref{fig:Ion-imaging-summary}
are interactive in some versions of this manuscript. Interactive versions,
and source data, are also available at \href{http://dx.doi.org/10.6084/m9.figshare.106343}{http://dx.doi.org/10.6084/m9.figshare.106343}.
}

The overall 3D photoelectron distribution obtained (integrated over
$t$) is shown in figure \ref{fig:Electron-imaging-summary}.  A
few of the major features in the dataset are labelled. Of particular
note is the signal along the laser axis leading to blurring in this
direction. Although the interaction region along the laser axis is
minimized by use of a pseudo 1D molecular beam, as discussed in section
\ref{sub:Coincidence-imaging-spectrometer}, ionization of background
gas along the laser axis cannot be prevented. However, in a coincidence
measurement this background signal can be removed at the analysis
stage by filtering the electron data for coincidences with parent
ions in the central ion spot (see figure \ref{fig:Ion-imaging-summary}
for the corresponding ion distribution). A slight up-down asymmetry
is visible in the images; the photoelectrons in the upper hemisphere
are those which initially have trajectories away from the detector,
and therefore spend longer in the interaction region as they must
be turned-around by the extraction fields. The asymmetry is therefore
ascribed to a combination of possible effects: slight inhomogeneities
in the extraction fields, especially near the grids, and perturbation
by weak external fields. Such effects could perturb the photoelectrons,
and would become more significant for longer interaction time-scales.
A central artefact is present around $(V_{x},V_{y})=0$ and to long
$ToF$s, the source of this artefact is unknown, but such structure
could arise from the creation of metastable autoioinzing states with
ns lifetimes (e.g. high-lying Rydberg states \cite{Chupka1993,Cockett2005}).
The strong probe-only signal, which appears in a narrow energy band
$E<0.3$~eV, is clearly visible near the centre of the distribution
and peaked along the laser polarisation axis. For $E>0.3$~eV all
electrons are from the pump-probe signal and, due to the rapid dynamics
resulting in a rapid shift of the IP (see figure \ref{fig:TRPES}),
appear quite diffuse in the 3D representation with no obvious angular
dependence in the $t$ integrated data. 

Figure \ref{fig:Ion-imaging-summary} gives an overview of the complete
ion data obtained (again integrated over all delays $t$), showing the full
3D map and projections onto 2D planes for (a) the parent ion mass
region around 54 a.m.u. and (c) the fragment region around 39~a.m.u.
The central panel (b) shows the calibrated mass spectrum obtained
by integrating over $(x,y)$, excluding the background gas signal
($x<4000$). The 3D ion map provides much more direct
information than the equivalent mapping of the electron data, and
from the figure various key features are immediately visible: (1)
the main parent ion feature, localised in $(x,y,ToF)$ and intense;
(2) satellite features to the parent ion, assigned as isotopes to
higher mass (parent+1 and parent+2) and hydrogen loss channels to
lower mass (parent-1, parent-2 and parent-3); (3) weak fragment channels
appearing at shorter $ToF$, assigned as fragmentation of the parent
ion; (4) a stripe of signal to the edge of the detector, arising
from ionization of background gas, i.e. molecules not entrained in
the molecular beam which have zero net translational velocity, so
appear along the laser propagation axis. The benefits of $(x,y)$
sensitivity to the mass spectrum are immediately obvious: the background
signal is easily gated out of the analysis. Mass resolution of $\ll1$~a.m.u.
is readily obtained under these experimental conditions (ion extraction
pulse of 550~V, acceleration voltage of 150~V, mass range 0~-~230~a.m.u.),
and butadiene isotope features are cleanly resolved in the mass spectrum.

Figure \ref{fig:Ion-imaging-summary}(a), and the inset to \ref{fig:Ion-imaging-summary}(b),
shows an expanded view of the parent ion and nearby features. Underlying
the main feature, and centred on it, is a diffuse signal with the
same $ToF$. This feature is approximately 3 orders of magnitude weaker
than the main feature. This diffuse feature is assigned to a combination
of (1) parent ions which fragment after extraction from the ionization
region (i.e. within the flight tube) and (2) parent ion formed from
dissociation of clusters within the ionization region. In both cases
there would be a kinetic energy release relative to the direct parent
ion signal, which would lead to the broadened, isotropic feature observed.
The dimer peak at 108 a.m.u. was also observed (not shown in figure
\ref{fig:Ion-imaging-summary}(b), normalized intensity 2x10$^{-4}$),
and dimer-1 and dimer+1 features were also just visible (normalized
intensities $<$5x10$^{-5}$), confirming the presence of dimers in
the molecular beam. In the work discussed herein no attempt was made
to investigate the diffuse feature further by, for instance, varying
the gas mixture or pulsed valve timing. The diffuse nature of this
feature, relative to the main parent-ion feature, means that most
of these ions can be gated out of the analysis, while those that remain
under the main feature hardly contribute to the total signal.

Further weak, somewhat diffuse and isotropic features are observed
at earlier $ToF$ than the parent ion. These features are assigned
to hydrogen loss channels, resulting in the species $C_{4}H_{5}^{+}$,
$C_{4}H_{4}^{+}$ and $C_{4}H_{3}^{+}$. These channels are 2~-~4
orders of magnitude weaker than the main parent ion signal. They show
some kinetic energy release, resulting in their diffuse appearance
in the imaging data and a broadening of the peaks in the $(x,y)$
integrated data, and this broadening in the $ToF$ clearly increases
with the number of $H$-atoms lost (see inset of figure \ref{fig:Ion-imaging-summary}(b))
. The isotropic nature of these distributions indicates the slow release
of the fragment, relative to the timescale of molecular rotations
($\sim$10s of picoseconds), as would be anticipated from the complex
dissociation dynamics. 

Figure \ref{fig:Ion-imaging-summary}(c) shows the imaging data for
the fragment region around 39~a.m.u. The main feature is assigned
to the methyl loss channel, resulting in a $C_{3}H_{3}^{+}$ fragment.
The higher mass fragment is assigned to the same channel for the +1
isotope, based on the intensity ratio, but could also contain contributions
from $C_{3}H_{4}^{+}$; the lower mass channel is assigned to $C_{3}H_{2}^{+}$.
As with the hydrogen loss features, these channels show isotropic
angular distributions, but with a larger spread of kinetic energy
release. At longer $ToF$s a diffuse tail is observed. The length
of the tail - stretching to the parent ion feature - indicates a relatively
long lifetime for the fragmenting complex, because fragmentation must
occur in all regions of the spectrometer to result in the large smearing
out of the fragment mass spectrum observed. Therefore the upper limit
for the timescale of fragmentation is the time taken for ion extraction
from the interaction and acceleration region of the spectrometer,
hence ns to $\mu$s timescales. Conversely, ions which appear at the
fragment $ToF$ must fragment before extraction, so cannot take longer
than a few ns to fragment. The large range in timescales here suggests
that multiple fragmentation pathways may play a role.

Also shown in the mass spectrum, figure \ref{fig:Ion-imaging-summary}(b),
is another fragment region centred at 28 a.m.u. This is assigned as
ethylene cation, $C_{2}H_{4}^{+}$. The satellite peaks, following
the same pattern as the methyl loss region, are assigned to the +1
isotope at 29~a.m.u., and further hydrogen loss resulting in $C_{2}H_{3}^{+}$
and $C_{2}H_{2}^{+}$. This region has no diffuse tail, indicating
more rapid fragmentation than for the methyl loss channels. No lower
masses were observed in the mass spectrum (down to the mass cut-off
at 3~a.m.u.), indicating that all lower mass fragments were produced
as neutrals.

At this level of representation, a few key aspects of the dynamics
can be discerned: the electron data appears relatively structureless,
hence spectrally broad and possibly varying rapidly as a function
of $t$; the ion data shows several fragmentation products, and some
limits to the timescales of fragmentation for different channels can
be intuited; the fragment angular distributions appear near isotropic,
although quantitative analysis is required to make this conclusion
definitive.

\subsection{Low-dimensionality mappings: time-resolved mass spectrum and photoelectron
spectrum\label{sub:Low-dimensionality-mappings}}

\begin{figure}
\begin{centering}
\includegraphics{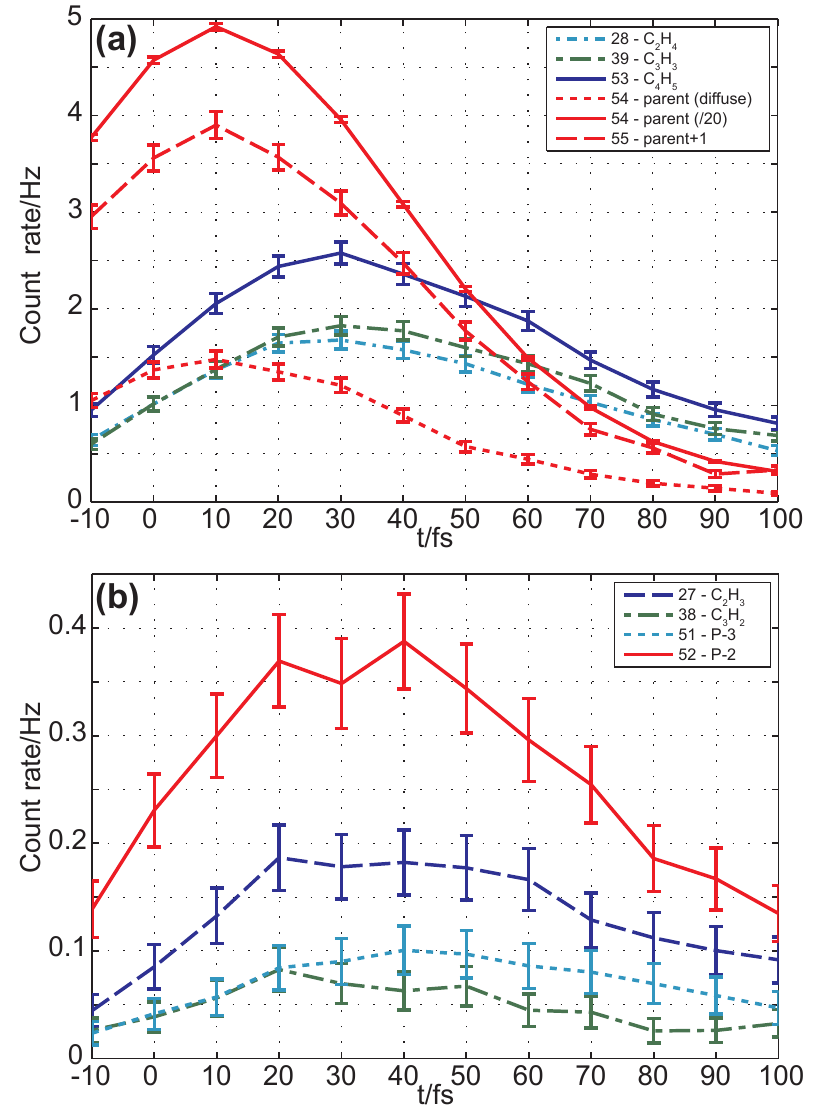}
\par\end{centering}

\caption{Time-resolved ion yields for a selection of the mass channels observed.
(a) Major ion channels. The parent ion signal is scaled down by a
factor of 20 for plotting purposes. (b) Minor channels. \label{fig:Time-resolved-ion-yields}}

\end{figure}

The next stage in analysis complexity is low-dimensionality mappings:
here we consider 2D mappings of the dynamics. First, the time-resolved
mass spectrum, the mapping of product yields vs. time for each mass
channel; second, the time-resolved photoelectron spectrum, the mapping
of electron yield vs. time for each kinetic energy channel. 

Figure \ref{fig:Time-resolved-ion-yields} shows the time-dependence
of a selection of the mass channels.  To obtain this representation,
the data was integrated over the $(x,y,ToF)$ coordinates for each
of the features of interest at each pump-probe delay $t$, and was
converted from raw counts to count rates to allow for the correct
weighting of the pump only and probe only signals, which could then
be subtracted from the full pump-probe signal.

The parent ion signal is observed to rise with the cross-correlation,
plateau and then fall with a Gaussian tail. The diffuse part of the
parent ion signal shows a very similar response, as do the isotope
peaks. The major hydrogen loss channel, $C_{4}H_{5}^{+}$, rises at
later $t$, and peaks around 20~fs after the parent ion. The minor
hydrogen loss channels, the parent-2 and parent-3 features shown in
figure \ref{fig:Time-resolved-ion-yields}(b), both peak at around
30~fs after the parent ion. Therefore, the data indicates that evolution
on the excited state of around 20~-~30~fs is required before these
fragmentation channels are open. Similarly, the fragment channels
assigned to $C_{3}H_{3}^{+}$ and $C_{2}H_{4}^{+}$ show peaks around
20~fs after the parent ion. In all cases the peak shapes are similar,
with non-Gaussian tails. Assuming that fragmentation only occurs on
the $D_{1}$ surface, the timescales here indicate that direct ionization
to $D_{1}$ is possible very rapidly, and indicates a significant
lowering of the vertical IP from the Franck-Condon region. At the
ground state equilibrium geometry $D_{0}$ and $D_{1}$ lie at 9.07~eV
and 11.39~eV respectively \cite{White1974}, so are separated by
$\sim$2.3~eV, with $D_{1}$ lying 0.53~eV above the available 1+1$'$
photon energy of 10.86~eV.  Hence the observed dynamics suggest
that the vertical IP to $D_{1}$ falls by $\sim$0.5~eV in $\sim$20~fs,
and this inferred drop is very similar to the shift observed in the
photoelectron signal, discussed below. Another possibility is that
the cross-section for absorption of a second probe photon increases
dramatically over the first 20~fs of the dynamics. Such 1+2$'$ processes
would provide 15.52~eV of energy, allowing population of several
higher-lying cationic states \cite{Dannacher1980}, either via direct
2-photon absorption, or sequential absorption via $D_{0}$.

The low laser fluences used experimentally suggest that all observed
signals arise from 1+1$'$ pump-probe processes, conversely the known
appearance energies of the fragments ($>$11.3~eV \cite{Dannacher1980,Fang2011})
suggest 1+2$'$ processes, may be energetically required for fragmentation
to occur.%
\footnote{Experimentally these studies could only probe the ground state minimum
geometry, so do not rule out the appearance of fragments at lower
energies as a function of nuclear coordinates, which would only be
limited by the asymptotic energy of the fragments. However, the energetics
of the dissociation pathways, including transition states and final
products, were also considered computationally in ref. \cite{Fang2011},
and the theoretical results also indicate that the observed fragments
are not energetically accessible via 1+1$'$ processes.} In order to check more carefully for 1+2$'$ processes the time-resolved
ion yields were also extracted in coincidence with photoelectrons
of energy 0~-~1.8~eV. This produced essentially the same traces
as shown in figure \ref{fig:Time-resolved-ion-yields}, except slightly
noisier. This does not completely rule out 1+2$'$ processes, which could
also produce photoelectrons in this energy region by population of
cation states with internal energies of $>$13.72~eV, or via the
sequential process of ionization and subsequent excitation of the
cation. A careful probe power study would be required to firmly answer
this question experimentally, but this remains for future work. 

In summary, the fragment yields peak around 20~-~30~fs after the
parent ion, indicating rapid dynamics lead to the opening of the observed
fragmentation channels. Additional filtering of the data for coincidences
with electrons in the 1-photon ionization region suggested, but cannot rigorously
confirm, that there is no significant 2-photon ionization contribution.
In a future publication, an extended analysis of the time-resolved
ion data will be made in order to examine the kinetic energy release
spectra of each fragment. Combined with the available fragment energetics
data this information may be sufficient to accurately determine mechanistic
details of the various dissociation pathways.

\begin{figure}
\begin{centering}
\includegraphics{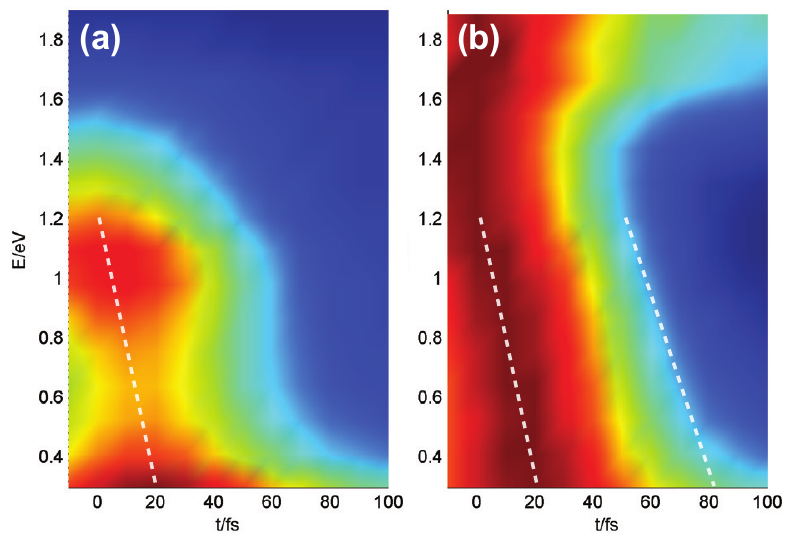}
\par\end{centering}

\caption{TRPES mapping. (a) Raw counts per $(E,\, t$) bin. The dashed line
follows the tilt of the band maximum, highlighting the chirp of the
signal, with gradient $\sim$20~fs/eV. (b) As (a), but normalized
to the maximum counts for each $E$ bin in order to show the temporal
behaviour at each energy independent of total counts. The second dashed
line follows the low edge of the signal, showing the slight broadening
as a function of energy, with gradient $\sim$30~fs/eV.\label{fig:TRPES}}

\end{figure}

We next consider the $(E,t)$ mapping of the electron data, the TRPES,
shown in figure \ref{fig:TRPES}. The TRPES shows an energetically
broad photoelectron band up to the 1-photon cut-off at $\sim$1.8~eV.
The low-energy cut-off at $\sim$0.3~eV corresponds to the region
of probe-only signal. The main feature at $\sim$1.2~eV follows the
cross-correlation of the pump and probe pulses ($\tau_{xc}\sim$60~fs
FWHM); outside of the cross-correlation region the signal rapidly
moves to lower energies, with a broad stripe visible in the data.
This stripe is slightly sloped or chirped (this is especially clear
in the energy normalized representation in figure \ref{fig:TRPES}(b)),
with the onset time varying as a function of energy. The chirp appears
close to linear over the observed energy region, with a gradient of
20$\pm$10~fs/eV. The temporal width of the band increases slightly
as a function of energy (shown by the second dashed line in figure
\ref{fig:TRPES}(b)), and the trailing edge of the band follows a
gradient of approximately 30$\pm$10~fs/eV, although there is increased,
non-linear, broadening below $\sim$0.4~eV. For the linear region
the broadening of the band is therefore $\sim$10~fs/eV; a higher
temporal resolution measurement - smaller $\Delta t$ - would provide
a more accurate figure. Apart from the slope of the band there is
little structure observed, and most of the population rapidly leaves
the observation window of the measurement (on the order of $\tau_{xc}$,
$\sim$60~fs). This data is very similar to previous TRPES studies
\cite{bogusNatChem}, but with shorter $\tau_{xc}$ and obtained for
only a small set of delays in order to ensure good statistics for
the TRPADs. Although the clipping of the data along the temporal coordinate
makes rigorous determination of the lineshapes impossible for the
peak of the signal in this dataset, based on the previous TRPES data
the temporal profile of the signal is assumed to be approximately
Gaussian for most, if not all, energies (see also the line-outs in
figure \ref{fig:TRPADs3D}), with a small (few percent of the total
counts) non-Gaussian tail outside of $\tau_{xc}$.

In terms of the dynamics, the delayed onset as a function of energy
is phenomenologically consistent with the picture of fast wavepacket
propagation on a steep potential energy surface. Such motion would
map primarily to the vertical IP (assuming that the excited state
and ionic ground state potential energy surfaces are not topologically
identical), hence the kinetic energy of the observed photoelectrons.
The TRPES provides information on the speed of the IP change, and
is consistent with the appearance of fragmentation channels with a
delayed onset which require a significant drop in the vertical IP
to $D_{1}$ relative to the Franck-Condon region. Conversely, wavepacket
dynamics which cause little change to the IP would be responsible
for signal observed in a given energy region at long delays, arising
from population which remains within the observation window of the
measurement (although such population may still trace complex trajectories
in energy space). Since the signal in all regions outside of the cross-correlation
time-scale is small, it is clear that wavepacket motion along these
coordinates is a very minor contribution to the dynamics. The data
also shows that the observed signal stays near Gaussian for all energy slices, again
indicating that there is little dispersion of the wavepacket. 

In broad terms, the ion and electron 2D data give some insight into
the wavepacket dynamics, with the general characteristics apparent
and some clocking of this motion possible. The observed energetic
shift in the photoelectron signal is consistent with the vertical
IP drop required to access $D_{1}$, hence observe fragmentation,
as already inferred from the ion data. In order to assemble a more
detailed picture we next consider the additional information available
from the TRPADs.

\subsection{High-dimensionality mappings: time-resolved photoelectron angular
distributions\label{sub:TRPADs_results}}

\begin{figure*}
\begin{centering}
\includegraphics{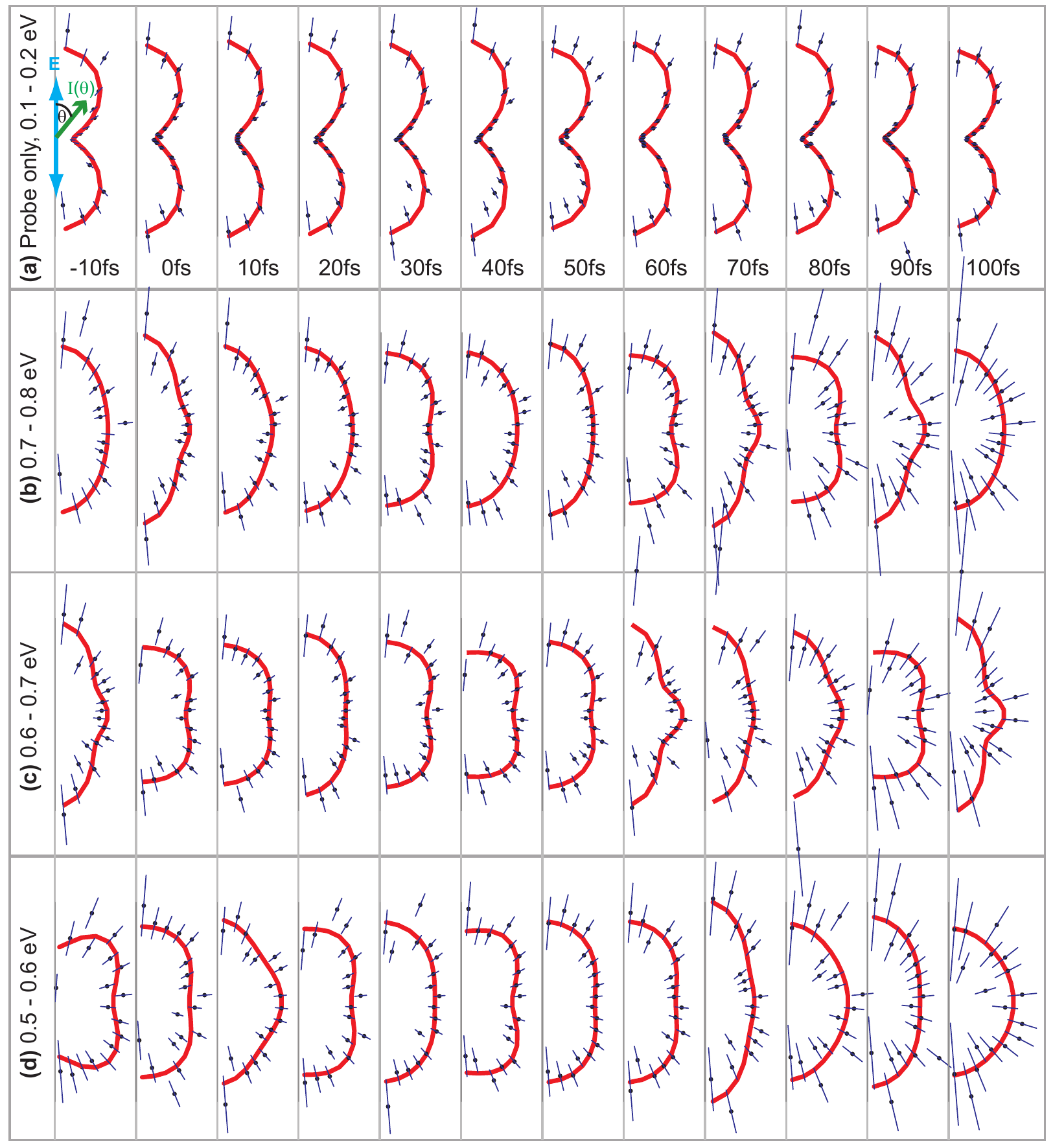}
\par\end{centering}

\caption{TRPADs, $I(\theta;t,E)$, extracted from the dataset as detailed in
the main text. (a) Probe only data, 0.1~-~0.2~eV. (b) - (d) Pump-probe data for 0.7~-0.8~eV, 0.6~-~0.7~eV and 0.5~-~0.6~eV respectively. Data points are shown with statistical error bars,
solid line shows fits to eqn. \ref{eq:IBlm} with $L=0,\,2,\,4$ and
$M=0$. \label{fig:TRPADs2D}}

\end{figure*}

In order to extract TRPADs the data is calibrated and rebinned in
polar coordinates as detailed in section \ref{sub:Data-analysis-and-calib}
to give the intensity (counts) per 4D volume element $\Delta\phi\sin\theta\Delta\theta\Delta E\Delta t$,
denoted $I(\theta,\phi;\, E,\, t)$. The choice of binning is, naturally,
limited by the experimental time-steps and instrument resolution;
however coarser binning can be used in order to improve the statistics
at the loss of resolution. For the data presented here $\Delta t=10$~fs
and $\Delta E=0.1$~eV, and in all cases shown here the data was
also integrated over $\phi$ which, due to the cylindrical symmetry
of this experiment, results in no loss of information but does improve
the statistics per $\theta$ bin. To obtain the cleanest possible
TRPADs the upper hemisphere electrons were discarded at the cost of
a factor of two in counts. The data was also filtered for coincidences
with the main parent ion feature. Because the parent ion dominates
the signal, this filtering was not required to obtain TRPADs correlated
with a given ion channel (although, more generally, could be used
in this way), but did serve to remove all signal from background gas
and scattered light, resulting in cleaner TRPADs albeit at the cost
of total counts. Figure \ref{fig:TRPADs2D} shows an example of the
TRPADs obtained in this way for four energy slices. In this representation,
the areas of the TRPADs are normalised to unity in order to allow comparison
of the form of the PADs independent of total counts. The error bars
and scatter of the data points should therefore be used as a guide
to the statistical significance of the extracted PADs over the various
energy and time slices shown. The solid lines show a fit to an expansion
in spherical harmonics, as defined by eqn. \ref{eq:IBlm}. 

From the data it is immediately clear that the TRPADs exhibit complex
behaviour. The TRPADs change rapidly as a function of $t$, with changes
on the order of 10~-~20~fs apparent. Because the laser pulses used
in this work were around 35~-~40~fs in duration, there is already
significant temporal blurring in the measured data. Despite this,
the changes are still clear and unambiguous in the data. For comparison,
TRPADs extracted for the probe only background (figure \ref{fig:TRPADs2D}(a))
show no significant changes temporally beyond the signal noise, and
consequent statistical variation of the fit, signifying that no temporal
artefacts are present in the raw data or introduced via the data processing.
The scatter in the data is worst at the poles of the distribution
due to the $\sin\theta$ normalisation factor which serves to amplify
noise near the poles; conversely the equatorial region shows very
little variability. There is some asymmetry present in the data which
is ascribed to a combination of noise (scatter) and detector artefacts/inhomogeneities.
For this latter reason the form of the extracted PADs (as defined
by the fitted $\beta_{LM}$ parameters) may not be highly accurate,
but the results do have high precision and reproducibility as shown
by figure \ref{fig:TRPADs2D}(a), so any relative temporal changes
observed should be reliable and robust within the statistical uncertainty.

\begin{figure*}
\begin{centering}
\includegraphics{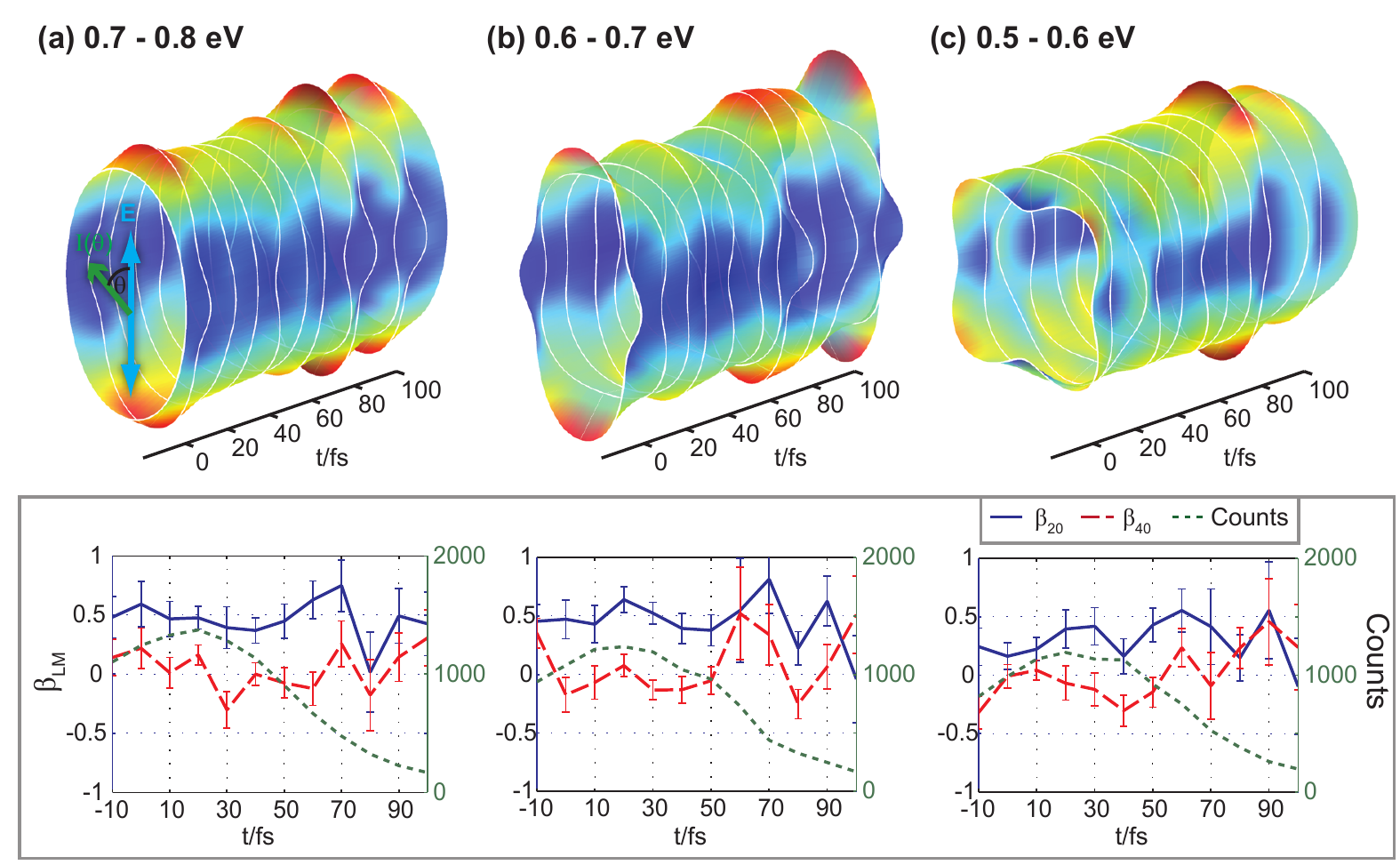}
\par\end{centering}

\caption{Fitted TRPADs, $I(\theta;t,E)$, as per figure \ref{fig:TRPADs2D},
represented as polar surface plots. White lines show the fits at 10~fs
intervals, the surface and colour map interpolate between these discrete
measurements. Lower panels show the fitted $\beta_{20}(t)$ and $\beta_{40}(t)$,
as well as the counts, for each energy region.\label{fig:TRPADs3D}
}

\end{figure*}

One challenge of high-dimensionality datasets is the presentation
and/or reduction of the data to a more tractable form to allow for
pattern recognition at a phenomenological or quantitative level. For
the TRPADs one can reduce the full dataset to the fitted TRPADs, which
can then be represented as $(E,I(\theta))$ or $(t,I(\theta))$ surfaces,
or maps of $\beta_{LM}(E,t)$. Figure \ref{fig:TRPADs3D} shows examples
of the TRPADs represented as $(t,I(\theta))$ surfaces (in polar space)
for the same three energy slices as figure \ref{fig:TRPADs2D}(b)-(d),
and the associated $\beta_{LM}(t)$ parameters are plotted in the
insets with error bars. Such representation, as a function of energy
or time, readily allows for comparison of the evolution of the shape
of the PADs, although visual information about the quality of the
raw data and fit fidelity are lost, so care must be exercised when
drawing conclusions from such maps. For the TRPADs shown in figure
\ref{fig:TRPADs3D}, the comments pertaining to figure \ref{fig:TRPADs2D}
can be reiterated, namely that the TRPADs for the different energy
regions are significantly different, and show complex temporal evolution.
More specifically, for the 0.5~-~0.6~eV region, the observed TRPADs
show a 4-lobed structure (significant $\beta_{4,0}$) at $t=-10$~fs,
which evolves to a 4-lobed structure with a different orientation
at $t=10$~fs, returns to near its initial shape at intermediate
$t$, and finally shows fast oscillations at $t>60$~fs, although
these later oscillations may not be reliable due to the low statistics
in this region. The colour mapping on the surface plots emphasises
the evolution of the signal intensity along the equator and at the
poles of the TRPADs; the corresponding $\beta_{LM}$ plots show that
both $\beta_{2,0}$ and $\beta_{4,0}$ change significantly as a function
of $t$, with $\beta_{2,0}$ displaying two peaks in the range $20\leq t\leq70$~fs.
The $\beta_{4,0}$ trace shows a minimum at $t=40$~fs and a maximum
at $t=60$~fs, which coincide with a mimimum and maximum in the $\beta_{2,0}$
trace, but at earlier times ($t<40$~fs) and later times ($t>60$~fs)
the behaviour does not appear to be directly correlated to the $\beta_{2,0}$
trace.

For the 0.6~-~0.7~eV energy slice the picture is quite different.
For the first time step the PAD again shows a 4-lobed structure, but
with intensity peaked at the poles and equator, as opposed to at 45$^{\circ}$
as per the 0.5~-~0.6~eV window. This corresponds to a large and
positive $\beta_{4,0}$, compared with a large and negative $\beta_{4,0}$
for the lower energy slice. The $\beta_{4,0}$ value goes negative,
with small magnitude, for the following time slices, and appears to
show a slight oscillation with minima at $t=0,\,40$~fs and a peak
at $t=20$~fs; the $\beta_{2,0}$ parameter shows a correlated oscillation
but centred around a mean value of $\sim0.5$. In the surface plots,
this oscillation appears as a slight breathing of the TRPADs, with
the largest changes at the poles. At later times, $t>50$~fs, more
complex behaviour is observed, with a significant beating around the
equator of the distributions as well as at the poles. 

For the 0.7~-~0.8~eV energy slice the behaviour is again different,
with much less variability in the observed TRPADs over the peak of
the signal. The $\beta_{2,0}$ value decreases gradually from the
local maxima at $t=0$~fs until $t=50$~fs, then increases gradually
to $t=70$~fs. The $\beta_{4,0}$ trace shows correlated local maxima
at $t=0,\,70$~fs, but much more variability between these peaks.
As was the case for the lower energy slices, the data at long delays
shows significant scatter and has low statistics, so should be treated
with care and carefully compared with the plots showing angular data
points (figure \ref{fig:TRPADs2D}) before drawing firm conclusions
as to the veracity of the temporal evolution observed in this region.
However, the fact that the observed $\beta_{2,0}$ dips at the same
time ($t=80$~fs) over all three energy slices, which were analysed
independently, suggests this behaviour is genuine and not the result
of random noise. In terms of the form of the TRPADs, the oscillations
in the $\beta_{L,M}$ describe significant oscillations which include
large changes to the photoelectron flux around the equator of the
distributions. Because the equatorial region is statistically more
reliable than the noisier poles this again suggests that these observations
are genuine.

\begin{figure*}
\begin{centering}
\includegraphics{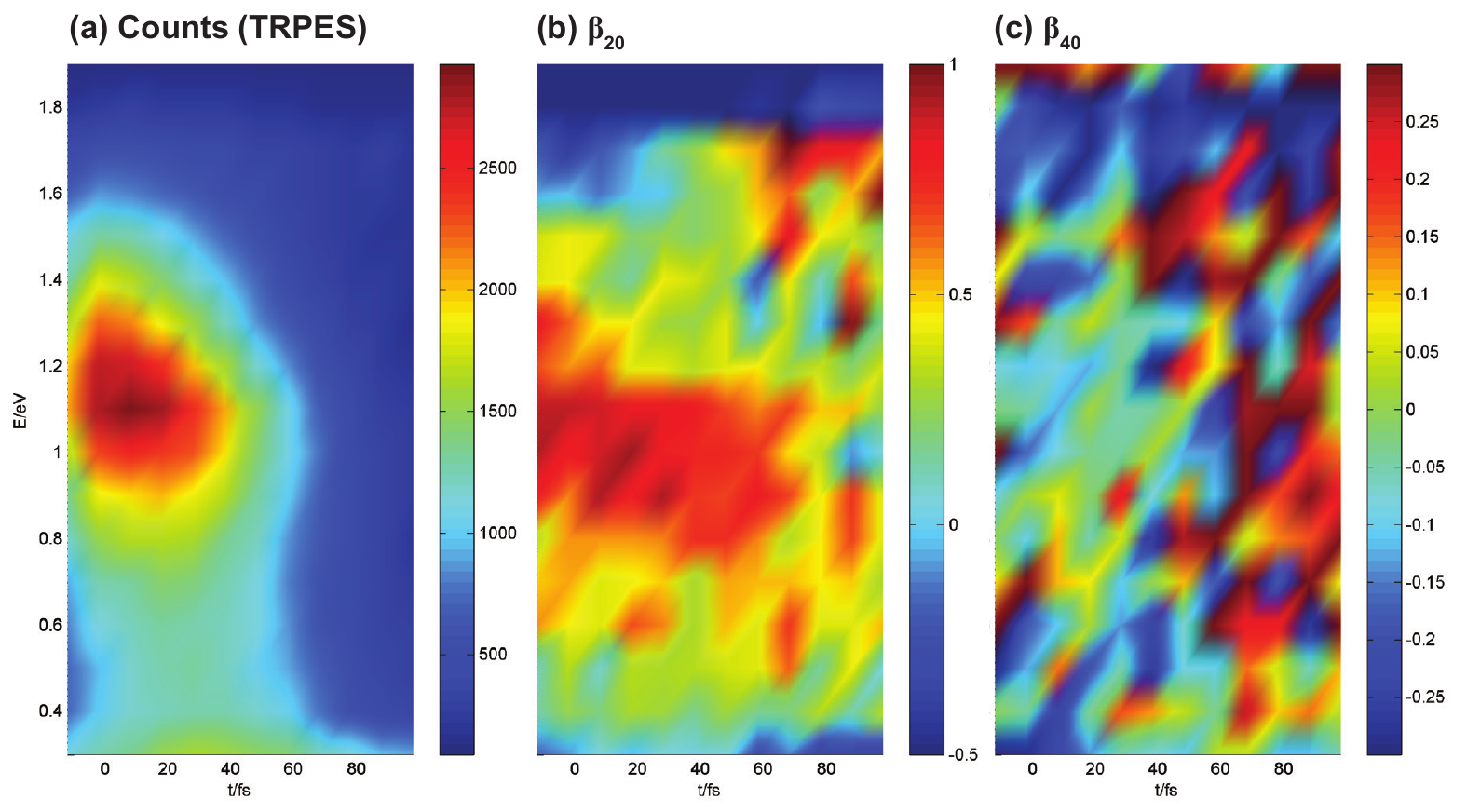}
\par\end{centering}

\caption{Energy-time maps of the total counts (TRPES) and extracted $\beta_{LM}$
parameters.\label{fig:BLM-maps}}

\end{figure*}

Figure \ref{fig:BLM-maps} shows maps of the $\beta_{LM}$ parameters
for all energy and time slices, along with the TRPES. This representation
of the TRPADs is essentially one step further removed from the raw
data, so again should be used in concert with plots showing the raw
data before drawing firm conclusions, but also provides the most reduced
and tractable form of the measurement. In this format, the 3 energy
regions discussed in detail in the preceding can be readily compared.
It is clear that the higher energy region correlates to larger $\beta_{2,0}$
values, which change little over the main part of the signal, while
the lower energy region shows marked oscillations; similar oscillations
extend to the lower energy slices, with peaks around $t=20,\,70$~fs.
At later times, $t>60$~fs, the trend is for reduced magnitude $\beta_{2,0}$,
particular over the main signal in the 0.9~-~1.3~eV region. These
times are outside of the cross-correlation of the laser pulses, so
should be representative of only the parts of the excited state wavepacket
which move orthogonal to the steep gradients on the potential energy
surface responsible for the speed of the dynamics and the rapid IP
change observed (i.e. population which remains in a given region of
configuration space for a longer time than the main part of the wavepacket).
The $\beta_{4,0}$ map shows much more oscillatory behaviour, which
appears to show no obvious correlation with energy or time, except
for the trend towards larger positive values at $t\geq60\, fs$. For
higher time resolution data it would be feasible to Fourier Transform
this data to extract the frequency content, but for the dataset shown
here the limited number of delays results in only a crude frequency
spectrum of little utility.

In summary, the TRPADs presented here contain a plethora of information,
and show very complex temporal evolution, in contrast to the TRPES
which provides little information on the temporal evolution of a given
energy slice over the probe pulse envelope, nor provides an observable
as sensitive to the underlying dynamics. The challenge for the experimentalist
is to determine whether the richness of the measured TRPADs can be
interpreted in terms of the underlying dynamics without recourse to
detailed theoretical treatments, that is to say without \emph{a priori}
knowledge of the underlying molecular dynamics or a full \emph{ab
initio} treatment of the dynamics and ionization. These points are
discussed further below (section \ref{sub:Discussion}).

\subsection{High-dimensionality mappings: correlated observables}

\begin{figure}
\begin{centering}
\includegraphics{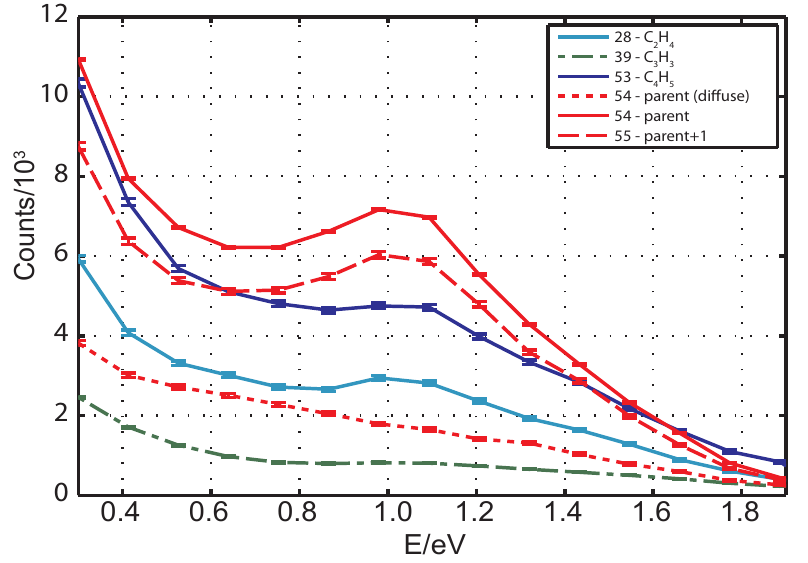}
\par\end{centering}

\caption{Time-integrated photoelectron spectra correlated with the major mass
channels (as shown in figure \ref{fig:Time-resolved-ion-yields}(a)).
The parent ion correlated spectrum is scaled down by a factor of 20
for plotting purposes.\label{fig:Time-integrated-photoelectron-spectra}}

\end{figure}

The advantages of measuring in coincidence have already been discussed
in terms of removal of background or other unwanted signals from the
data hypercube. Naturally a further advantage is the ability to look
for ion-electron correlations, and retrieve minor channels which would
otherwise be inaccessible; for instance, the photoelectron spectra
of the fragment channels, which would not be distinguishable in a
non-coincidence measurement.

Figure \ref{fig:Time-integrated-photoelectron-spectra} shows the
photoelectron spectra (time-integrated) for the major ion channels
as shown in figure \ref{fig:Time-resolved-ion-yields}. The parent
ion correlated spectra shows the main cross-correlation feature around
1~eV, and a rise towards lower energy, as already seen in the TRPES
mapping (figure \ref{fig:TRPES}). As expected, the parent+1 channel
has an identical spectrum. However, the diffuse part of the parent
ion signal has a very different spectrum, consistent with a different
ionization process. Similarly, the fragmentation channels show different
spectra; in all cases the spectrum is broad, and there is a drop or
even disappearance of the 1~eV peak seen in the parent ion channel.
These spectra therefore provide additional information towards understanding
the excited state dynamics, with the spectra providing a fingerprint
of different ionization channels. Here, the suppression of the cross-correlation
peak in the fragment channels is consistent with the delayed opening
of these channels, as observed in the time-resolved ion yields. 

Furthermore, the similarity between the channels suggests ionization
occurs from a similar region of configuration space in all cases.
This is consistent with the similarity of the rise times and fall
times observed in the time-resolved fragment yields, which intuitively
suggests only a single dynamical pathway, shared by all the fragment
channels, hinting at localization of the excited state wavepacket
in configuration space. The similarity of the time-scales of the fragment
channels, considered in light of the similarities in the correlated
photoelectron spectra, therefore suggests that the fragmentation pathway
of the ion is very sensitive to the form of the wavepacket at the
time of ionization and, possibly, bifurcation on the ionic state(s)
may lead to channels with apparently similar temporal response, but
quite different fragmentation products. This is consistent with the
picture of rapid wavepacket motion, with little dispersion, on the
excited state and additionally indicates complex dissociation dynamics
on the ionic surfaces. In terms of the experimentally accessed ionization
pathways, the observation of similar photoelectron spectra for the
fragment channels is consistent with the possibility of sequential
1+2$'$ processes (ionization to $D_{0}$ followed by absorption of a
second probe photon) as discussed in section \ref{sub:Low-dimensionality-mappings},
but suggests that direct 1+2$'$ ionization to higher lying cationic
states, and 1+1$'$ ionization to $D_{1}$, are unlikely channels. Direct
ionization to excited ionic states would be expected to correlate
generally with different spectra, although it is always possible that
the spectra would not be significantly different - especially given
the diffuse, unstructured nature of the photoelectron bands of butadiene
- so this is not a rigorous conclusion.

Various other correlated mappings are available from the data, and
will be explored in a future publication. For example, the fragment
correlated TRPES and TRPADs, although for weak channels the latter
is very demanding statistically. For photodissociation studies energy
correlation spectra - maps of electon vs. ion kinetic energy - are
also useful \cite{Davies1999}.

\subsection{Discussion - mapping dynamics with multi-dimensional measurements\label{sub:Discussion}}

The data presented herein indicates that rapid and complex dynamics
are present in butadiene, consistent with earlier experimental \cite{Fuss2001,Assenmacher2001,bogusNatChem}
and theoretical \cite{Garavelli2001,Levine2009} works. The benefit
of a multidimensional dataset is clear, with complementary information
available from the coincident electron and ion data. The overall shape of the
dynamics, that of a near-Gaussian wavepacket which moves along steep
gradients on the potential energy surface which are strongly correlated
with the vertical IP, and shows little dispersion along other coordinates,
emerges rapidly from the data. This picture fits both the chirped
TRPES data and the delayed-onset of fragmentation channels which are
assumed to be correlated with ionization to $D_{1}$. The observation
of TRPADs which show very rich behaviour on rapid timescales ($<$20~fs)
is striking, and indicates the sensitivity of the TRPADs to the dynamics
under study.

However, more specific mechanistic details, such as the nuclear motions
involved or the mapping of these motions onto the TRPADs, do not seem
to be forthcoming. This is in contrast to simpler cases, such as previous
work on the $NO$ dimer \cite{Gessner2006} or $CS_{2}$ \cite{Bisgaard2009,Hockett2011},
in which the limited dimensionality of the problem enabled a more
direct empirical approach, coupled with symmetry-based modelling to
understand the measurements in more detail. Understanding the information
conveyed by the TRPADs from larger systems therefore remains a significant
challenge.

In the case of butadiene, detailed \emph{ab initio} dynamics calculations
do exist \cite{Levine2009}, and further studies are ongoing \cite{schuurman2012}.
It is therefore very tempting to compare the measurements to these
calculations, despite the fact that these calculations do not, so
far, explicitly include the ionization matrix elements. For instance,
figures 5 and 6 of ref. \cite{Levine2009} show the behaviour and
timescales associated with the bond alternation coordinate (a motion
involving bonds along the carbon backbone streching and compressing)
and the out-of-plane twisting motions (twisting about the central
CC bond and twisting of the terminal methyl groups). The bond alternation
coordinate shows oscillations with a period of $\sim$20~fs, and
large amplitude motion with changes up to 0.4~\AA. The twisting coordinates
show rapid changes over $\sim40$~fs, followed by almost constant
bond angles, with some fluctuations, for the remaining 160~fs of
the calculations. These motions are very suggestive of the oscillations
observed in $\beta_{2,0}(t)$ for some of the low-energy slices, and
the gradual rise over $\sim$60~fs observed in $\beta_{4,0}(t)$
over all photoelectron energies. However, from the data presented
in ref. \cite{Levine2009} it is not apparent how these motions map
to the IP, which is a very significant factor in drawing these conclusions
more firmly. As shown in ref. \cite{Arasaki2010} such mappings may
be very complex, and difficult to determine from anything but a full
calculation including the ionization matrix elements. 

Another conclusion from the dynamics calculations, which is likely
to be more robust, is the agreement between the timescale of the population
dynamics and the gross changes in both $\beta_{2,0}(t)$ and $\beta_{4,0}(t)$
at all energies for $t\geq60$~fs. Figure 4 of ref. \cite{Levine2009}
shows the populations of the bright and dark adiabatic states, defined
there as $S_{1}$ and $S_{2}$ respectively. The $S_{1}$ population
dips almost immediately from its initial value at $t=0$, with almost
equal populations of $S_{1}$ and $S_{2}$ at $t\approx25$~fs, the
$S_{1}$ population then grows again and peaks at around 50~fs, while
the $S_{2}$ population drops rapidly. Unsurprisingly, given the similarity
of the timescales to those mentioned above, the non-adiabatic coupling
between these states is primarily mediated by the out-of-plane twisting
motions. In terms of the electronic character of the states, the
early time dynamics is best considered in a diabatic picture, with
the bright state $^{1}B_{u}$ character maintained as population switches
from the $S_{1}$ to the $S_{2}$ surface at early times \cite{Levine2009},
while over the longer timescale the electronic character becomes mixed
by the out-of-plane twisting, and population is transferred non-adiabatically
between $S_{2}$ and $S_{1}$. 

Given that the adiabatic states are of different electronic character
one would expect a different PAD from each state; additionally, interferences
between photoelectrons from the two states (different parts of a split
wavepacket) may also play a significant role in the observable if
electrons of the same energy are created from the two states, and
this could lead to more rapid modulations in the form of the TRPADs
(this is essentially the concept of the two-state model previously
applied to $CS_{2}$ \cite{Hockett2011}), particularly in the case
of symmetry-breaking leading to mixing of the adiabatic electronic
state characters. With this in mind, it appears that the (adiabatic)
population dynamics may have a more direct link to the observed TRPADs,
with the changes at $t\geq60$~fs corresponding to the region where
most of the population is on a single adiabatic state, and the molecular
geometry, hence electronic character, has stabilised after the initial,
rapid, out-of-plane twisting. Although this picture seems reasonable,
without further modelling or calculations these conclusions remain
somewhat tentative.

The challenge for the experimentalist therefore remains: to determine
how the TRPADs, and other correlated observables, can be interpreted
without recourse to detailed theoretical treatments, hence
without \emph{a priori} knowledge of the underlying molecular dynamics
or a full \emph{ab initio} treatment of the dynamics and ionization.
While such computational approaches are very powerful, they are also
time consuming, difficult, and only computationally tractable for
small molecules. Naturally the same applies to a purely experimental
approach to the TRPES and TRMS data, but typically the lower dimensionality
of this data lends itself to only broad interpretations, such as the
presence of band switching \cite{Blanchet1999} or quantum beats,
which may fully describe the wavepacket dynamics in small systems
\cite{Fischer1996} or arise from only the recurrent parts of the
wavepacket \cite{Bisgaard2009,Hockett2011} in more complex cases.
Moreover, TRPES data is routinely interpreted via fitting of temporal
profiles in what is essentially a principle component or single-value
decomposition analysis \cite{Wu2011a,VanWilderen2011}. While such
analysis provides a quantitative breakdown of the data into spectral
and temporal functions, it makes many assumptions regarding the underlying
dynamics (vis. that the observables equate directly to state populations
in the standard interpretations, or at the very least behave phenomenologically
in a statistical manner) and cannot be used to treat more complex
dynamics, such as the kind observed here, without recourse to many
time constants. The TRPADs clearly convey more information on the
wavepacket dynamics, but the mapping from wavepacket to observable
is inherently complicated.

Aside from the TRPADs, the full imaging data provides further correlated
observables which may also be considered in light of the computational
results. In particular, the time-resolved ion yields and fragment-correlated
photoelectron spectra may provide a way of experimentally probing
the three $S_{1}\rightarrow S_{0}$ CIs discussed in ref. \cite{Levine2009},
which correspond to different molecular geometries. It is likely that
ionization from these different regions of configuration space (if
energetically possible in the limited observation window) would lead
to different photoelectron spectra and different fragmentation products
from $D_{1}$. Even in the case where direct ionization from these
regions is energetically forbidden, ionization of the parts of the
wavepacket on $S_{1}$ heading towards these distinct regions of configuration
space would also result in different dynamics on the ion surfaces,
hence possibly lead to different fragmentation products. Further
analysis of ion-electron correlations should present deeper insight
into the excited state dynamics than uncorrelated ion or electron
measurements alone.

\section{Conclusions \& future work}

The data presented here illustrates the richness of coincidence imaging
datasets, even in the case of very rapid and complex dynamics; it
also highlights the difficulty of interpretation of such data. In
this work, we have focussed on the TRPADs obtained in coincidence with
the parent ion channel, and made some tentative comparisons with theory.
In general the full 7D data provides the best hope of understanding the
excited state dynamics of molecules but, unsurprisingly, significant
effort is required to gain understanding directly from the data, or
with the aid of theory. Despite the depth of material presented here
there are still many avenues to explore in order to obtain more complete,
unified experimental and theoretical approaches to problems in excited
state dynamics.

In future work we hope to continue bridging this gap between experiment
and theory via a number of routes: by the more direct and explicit
comparison of the experimental data with dynamics calculations, and
the inclusion of observables in the calculations; by exploring further
the possibilities of modelling for a qualitative/semi-quantitative
treatment of the dynamics, expanding upon previous work which employed
a simple 2-state model \cite{Hockett2011}, possibly to include a
basic wavepacket treatment; through further analysis of the ion fragment
distributions, which have not yet been fully explored; by making further
experimental measurements for aligned systems and with different pump-probe
polarisation geometries in order to obtain more detailed TRPADs (in
both cases  more $L,\, M$ terms are allowed in the PADs), with harder
VUV probe photons to access a larger photoelectron energy range, and
with a stability enhanced set-up to allow for longer experimental
runs; by further use of fragment correlated TRPES/TRPADs, so far not
statistically feasible due to the dominance of the parent ion channel
and consequent low total counts in these fragment channels. This multi-dimensional
approach to a multi-dimensional challenge will hopefully prove fruitful
for a deeper understanding of excited state dynamics in polyatomic molecules and,
in particular, help to make the experimental measurement of such dynamics
a more routine and insightful tool.

\section{Acknowledgements}

We thank Owen Clarkin and Jochen Mikosch for general assistance with
the CIS instrument, Rune Lausten for assistance with the laser system
and optical set-up, Doug Moffat for assistance with the data acquisition
software and Denis Guay for assistance with vacuum hardware. We also
thank Andrey Boguslavskiy and Michael Schuurman for helpful discussions
on butadiene and associated dynamics calculations.

\bibliographystyle{tMOP}
\bibliography{butadiene_JMOpaper_refs_all_150113}

\begin{thebibliography}{72}
\providecommand{\natexlab}[1]{#1}

\bibitem[1]{Moshammer1994}
Moshammer, R.; Ullrich, J.; Unverzagt, M.; Schmidt, W.; Jardin, P.; Olson, R.;
  Mann, R.; D\"{o}rner, R.; Mergel, V.; Buck, U.; et~al. {Low-Energy Electrons
  and Their Dynamical Correlation with Recoil Ions for Single Ionization of
  Helium by Fast, Heavy-Ion Impact}.  {\em Physical Review Letters}  {\bf
  1994}, {\em 73} (25) (Dec.), 3371--3374.

\bibitem[2]{Moshammer1996}
Moshammer, R.; Unverzagt, M.; Schmitt, W.; Ullrich, J.; et~al. {A 4$\pi$
  recoil-ion electron momentum analyzer: a high-resolution “microscope” for
  the investigation of the dynamics of atomic, molecular and nuclear
  reactions}.  {\em Nuclear Instruments and Methods in Physics Research Section
  B: Beam Interactions with Materials and Atoms}  {\bf 1996}, {\em 108} (4)
  (Mar.), 425--445.

\bibitem[3]{Ullrich1997}
Ullrich, J.; Moshammer, R.; D\"{o}rner, R.; Jagutzki, O.; Mergel, V.;
  Schmidt-B\"{o}cking, H.; et~al. {Recoil-ion momentum spectroscopy}.  {\em
  Journal of Physics B: Atomic, Molecular and Optical Physics}  {\bf 1997},
  {\em 30} (13) (Jul.), 2917--2974.

\bibitem[4]{Dorner2000}
D\"{o}rner, R.; Mergel, V.; Jagutzki, O.; Spielberger, L.; Ullrich, J.;
  Moshammer, R.; et~al. {Cold Target Recoil Ion Momentum Spectroscopy: a
  ‘momentum microscope’ to view atomic collision dynamics}.  {\em Physics
  Reports}  {\bf 2000}, {\em 330} (2-3) (Jun.), 95--192.

\bibitem[5]{Czasch2005}
Czasch, A.; Schmidt, L.; Jahnke, T.; Weber, T.; Jagutzki, O.; Sch\"{o}ssler,
  S.; Sch\"{o}ffler, M.; D\"{o}rner, R.; et~al. {Photo induced multiple
  fragmentation of atoms and molecules: Dynamics of Coulombic many-particle
  systems studied with the COLTRIMS reaction microscope}.  {\em Physics Letters
  A}  {\bf 2005}, {\em 347} (1-3) (Nov.), 95--102.

\bibitem[6]{Landers2001}
Landers, A.; Weber, T.; Ali, I.; Cassimi, A.; Hattass, M.; Jagutzki, O.;
  Nauert, A.; Osipov, T.; Staudte, A.; Prior, M.; Schmidt-B\"{o}cking, H.;
  Cocke, C.; et~al. {Photoelectron Diffraction Mapping: Molecules Illuminated
  from Within}.  {\em Physical Review Letters}  {\bf 2001}, {\em 87} (1)
  (Jun.), 1--4.

\bibitem[7]{Akoury2007}
Akoury, D.; Kreidi, K.; Jahnke, T.; Weber, T.; Staudte, A.; Sch\"{o}ffler, M.;
  Neumann, N.; Titze, J.; Schmidt, L.P.H.; Czasch, A.; Jagutzki, O.; {Costa
  Fraga}, R.A.; Grisenti, R.E.; {D\'{\i}ez Mui\~{n}o}, R.; Cherepkov, N.A.;
  Semenov, S.K.; Ranitovic, P.; Cocke, C.L.; Osipov, T.; Adaniya, H.; Thompson,
  J.C.; Prior, M.H.; Belkacem, A.; Landers, A.L.; Schmidt-B\"{o}cking, H.;
  et~al. {The simplest double slit: interference and entanglement in double
  photoionization of H2.}.  {\em Science}  {\bf 2007}, {\em 318} (5852),
  949--952.

\bibitem[8]{Meckel2008}
Meckel, M.; Comtois, D.; Zeidler, D.; Staudte, A.; Pavicic, D.; Bandulet, H.C.;
  P\'{e}pin, H.; Kieffer, J.C.; D\"{o}rner, R.; Villeneuve, D.M.; et~al.
  {Laser-induced electron tunneling and diffraction.}.  {\em Science}  {\bf
  2008}, {\em 320} (5882) (Jun.), 1478--82.

\bibitem[9]{Eckle2008}
Eckle, P.; Pfeiffer, A.N.; Cirelli, C.; Staudte, A.; D\"{o}rner, R.; Muller,
  H.G.; B\"{u}ttiker, M.; et~al. {Attosecond ionization and tunneling delay
  time measurements in helium.}.  {\em Science (New York, N.Y.)}  {\bf 2008},
  {\em 322} (5907) (Dec.), 1525--9.

\bibitem[10]{Ueda2003}
Ueda, K. {High-resolution inner-shell spectroscopies of free atoms and
  molecules using soft-x-ray beamlines at the third-generation synchrotron
  radiation sources}.  {\em Journal of Physics B: Atomic, Molecular and Optical
  Physics}  {\bf 2003}, {\em 36} (4) (Feb.), R1--R47.

\bibitem[11]{Reid2012}
Reid, K.L. {Photoelectron angular distributions: developments in applications
  to isolated molecular systems}.  {\em Molecular Physics}  {\bf 2012}, {\em
  110} (3) (Feb.), 131--147.

\bibitem[12]{Davies1999}
Davies, J.A.; LeClaire, J.E.; Continetti, R.E.; et~al. {Femtosecond
  time-resolved photoelectron-photoion coincidence imaging studies of
  dissociation dynamics}.  {\em The Journal of Chemical Physics}  {\bf 1999},
  {\em 111} (1), 1.

\bibitem[13]{Lafosse2000}
Lafosse, A.; Lebech, M.; Brenot, J.; Guyon, P.; Jagutzki, O.; Spielberger, L.;
  Vervloet, M.; Houver, J.; et~al. {Vector Correlations in Dissociative
  Photoionization of Diatomic Molecules in the VUV Range: Strong Anisotropies
  in Electron Emission from Spatially Oriented NO Molecules}.  {\em Physical
  Review Letters}  {\bf 2000}, {\em 84} (26) (Jun.), 5987--5990.

\bibitem[14]{Continetti2001}
Continetti, R.E. Coincidence Spectroscopy.  {\em Annual Review of Physical
  Chemistry}  {\bf 2001}, {\em 52} (1), 165--192.

\bibitem[15]{Lebech2002}
Lebech, M.; Houver, J.C.; Dowek, D. {Ion-electron velocity vector correlations
  in dissociative photoionization of simple molecules using electrostatic
  lenses}.  {\em Review of Scientific Instruments}  {\bf 2002}, {\em 73} (4),
  1866.

\bibitem[16]{Gessner2006}
Gessner, O.; Lee, A.M.D.; Shaffer, J.P.; Reisler, H.; Levchenko, S.V.; Krylov,
  A.I.; Underwood, J.G.; Shi, H.; East, A.L.L.; Wardlaw, D.M.; Chrysostom,
  E.t.H.; Hayden, C.C.; et~al. Femtosecond Multidimensional Imaging of a
  Molecular Dissociation.  {\em Science}  {\bf 2006}, {\em 311} (5758),
  219--222.

\bibitem[17]{Vredenborg2008}
Vredenborg, A.; Roeterdink, W.G.; Janssen, M.H.M. {A photoelectron-photoion
  coincidence imaging apparatus for femtosecond time-resolved molecular
  dynamics with electron time-of-flight resolution of sigma=18 ps and energy
  resolution Delta E/E=3.5\%.}.  {\em The Review of Scientific Instruments}
  {\bf 2008}, {\em 79} (6) (Jun.), 063108.

\bibitem[18]{Bisgaard2009}
Bisgaard, C.Z.; Clarkin, O.J.; Wu, G.; Lee, A.M.D.; Gessner, O.; Hayden, C.C.;
  et~al. Time-Resolved Molecular Frame Dynamics of Fixed-in-Space CS2
  Molecules.  {\em Science}  {\bf 2009}, {\em 323} (5920), 1464--1468.

\bibitem[19]{Davies2000}
Davies, J.A.; Continetti, R.E.; Chandler, D.W.; et~al. {Femtosecond
  time-resolved photoelectron angular distributions probed during
  photodissociation of NO2.}.  {\em Physical Review Letters}  {\bf 2000}, {\em
  84} (26 Pt 1) (Jun.), 5983--6.

\bibitem[20]{Lee2007}
Lee, A.M.D. {Chemical Reaction Dynamics and Coincidence Imaging Spectroscopy}.
  Ph.D. thesis, Queen's University, 2007.

\bibitem[21]{Liu2011}
Liu, S.Y.; Alnama, K.; Matsumoto, J.; Nishizawa, K.; Kohguchi, H.; Lee, Y.P.;
  et~al. {He I ultraviolet photoelectron spectroscopy of benzene and pyridine
  in supersonic molecular beams using photoelectron imaging.}.  {\em The
  Journal of Physical Chemistry A}  {\bf 2011}, {\em 115} (14) (Apr.),
  2953--65.

\bibitem[22]{clarkin2012}
Clarkin, O.J. Chemical Reaction Dynamics at the Statistical Ensemble and
  Molecular Frame Limits. Ph.D. thesis, Queen's University, 2012.

\bibitem[23]{owenScattLight2012}
Clarkin, O.J.; Hockett, P.; Mikosch, J.; et~al. {\em In Preparation}  {\bf
  2013}.

\bibitem[24]{Rijs2004}
Rijs, A.; Janssen, M.; Chrysostom, E.; et~al. {Femtosecond Coincidence Imaging
  of Multichannel Multiphoton Dynamics}.  {\em Physical Review Letters}  {\bf
  2004}, {\em 92} (12) (Mar.), 1--4.

\bibitem[25]{Arasaki2000}
Arasaki, Y.; Takatsuka, K.; Wang, K.; et~al. Femtosecond energy- and
  angle-resolved photoelectron spectroscopy.  {\em The Journal of Chemical
  Physics}  {\bf 2000}, {\em 112} (20), 8871--8884.

\bibitem[26]{Seideman2002}
Seideman, T. {Time-resolved photoelectron angular distributions: concepts,
  applications, and directions.}.  {\em Annual Review of Physical Chemistry}
  {\bf 2002}, {\em 53} (Jan.), 41--65.

\bibitem[27]{Suzuki2003}
Suzuki, Y.; Stener, M.; Seideman, T. Multidimensional calculation of
  time-resolved photoelectron angular distributions: The internal conversion
  dynamics of pyrazine.  {\em The Journal of Chemical Physics}  {\bf 2003},
  {\em 118} (10), 4432--4443.

\bibitem[28]{Stolow2008}
Stolow, A.; Underwood, J.G., Time-Resolved Photoelectron Spectroscopy of
  Nonadiabatic Dynamics in Polyatomic Molecules,  {\em Advances in Chemical
  Physics }; Vol.  139, 2008; .

\bibitem[29]{Arasaki2010}
Arasaki, Y.; Takatsuka, K.; Wang, K.; et~al. {Time-resolved photoelectron
  spectroscopy of wavepackets through a conical intersection in NO2.}.  {\em
  The Journal of Chemical Physics}  {\bf 2010}, {\em 132} (12) (Mar.), 124307.

\bibitem[30]{Wu2011}
Wu, G.; Hockett, P.; Stolow, A. {Time-resolved photoelectron spectroscopy: from
  wavepackets to observables.}.  {\em Physical Chemistry Chemical Physics}
  {\bf 2011}, {\em 13} (41) (Nov.), 18447--67.

\bibitem[31]{Blanchet1999}
Blanchet, V.; Zgierski, M.Z.; Seideman, T.; et~al. {Discerning vibronic
  molecular dynamics using time-resolved photoelectron spectroscopy}.  {\em
  Nature}  {\bf 1999}, {\em 401}, 52--54.

\bibitem[32]{Blanchet2001}
Blanchet, V.; Zgierski, M.Z.; Stolow, A. Electronic continua in time-resolved
  photoelectron spectroscopy. I. Complementary ionization correlations.  {\em
  The Journal of Chemical Physics}  {\bf 2001}, {\em 114} (3), 1194--1205.

\bibitem[33]{Schmitt2001}
Schmitt, M.; Lochbrunner, S.; Shaffer, J.P.; Larsen, J.J.; Zgierski, M.Z.;
  et~al. Electronic continua in time-resolved photoelectron spectroscopy. II.
  Corresponding ionization correlations.  {\em The Journal of Chemical Physics}
   {\bf 2001}, {\em 114} (3), 1206--1213.

\bibitem[34]{Shapiro1999}
Shapiro, M.; Vrakking, M.J.J.; Stolow, A. {Nonadiabatic wave packet dynamics:
  Experiment and theory in IBr}.  {\em The Journal of Chemical Physics}  {\bf
  1999}, {\em 110} (5), 2465.

\bibitem[35]{Arasaki2000a}
Arasaki, Y.; Takatsuka, K.; Wang, K.; et~al. Probing wavepacket dynamics with
  femtosecond energy- and angle-resolved photoelectron spectroscopy.  {\em
  Journal of Electron Spectroscopy and Related Phenomena}  {\bf 2000}, {\em
  108} (1-3), 89 -- 98.

\bibitem[36]{Wollenhaupt2005}
Wollenhaupt, M.; Engel, V.; Baumert, T. {Femtosecond laser photoelectron
  spectroscopy on atoms and small molecules: prototype studies in quantum
  control.}.  {\em Annual Review of Physical Chemistry}  {\bf 2005}, {\em 56}
  (Jan.), 25--56.

\bibitem[37]{Hockett2011}
Hockett, P.; Bisgaard, C.Z.; Clarkin, O.J.; et~al. {Time-resolved imaging of
  purely valence-electron dynamics during a chemical reaction}.  {\em Nature
  Physics}  {\bf 2011}, {\em 7} (8) (Apr.), 612--615.

\bibitem[38]{bogusNatChem}
Boguslavskiy, A.E.; Schalk, O.; Schuurman, M.; Townsend, D.; et~al. Excited
  State Dynamics in the Smallest Polyene: Ultrafast Time-Resolved
  Photoelectron-Photoion Coincidence Spectroscopy of 1,3-Butadiene.  {\em In
  Preparation}  {\bf 2013}.

\bibitem[39]{Levine2009}
Levine, B.G.; Mart\'{\i}nez, T.J. {Ab initio multiple spawning dynamics of
  excited butadiene: role of charge transfer.}.  {\em The Journal of Physical
  Chemistry A}  {\bf 2009}, {\em 113} (46) (Nov.), 12815--24.

\bibitem[40]{Tao2011}
Tao, H.; Allison, T.K.; Wright, T.W.; Stooke, a.M.; Khurmi, C.; van Tilborg,
  J.; Liu, Y.; Falcone, R.W.; Belkacem, A.; et~al. {Ultrafast internal
  conversion in ethylene. I. The excited state lifetime.}.  {\em The Journal of
  Chemical Physics}  {\bf 2011}, {\em 134} (24) (Jun.), 244306.

\bibitem[41]{Leopold1984}
Leopold, D.G.; Pendley, R.D.; Roebber, J.L.; Hemley, R.J.; et~al. {Direct
  absorption spectroscopy of jet-cooled polyenes. II. The 1 1B+u←1 1A−g
  transitions of butadienes and hexatrienes}.  {\em The Journal of Chemical
  Physics}  {\bf 1984}, {\em 81} (10) (Nov.), 4218.

\bibitem[42]{Krawczyk2000}
Krawczyk, R.; Malsch, K.; Hohlneicher, G. {11Bu–21Ag conical intersection in
  trans-butadiene: ultrafast dynamics and optical spectra}.  {\em Chemical
  Physics}  {\bf 2000},  (April), 535--541.

\bibitem[43]{Assenmacher2001}
Assenmacher, F.; Gutmann, M.; Hohlneicher, G.; Stert, V.; et~al. {Ultrafast
  dynamics of the 11Bu-state of 1,3-butadiene after excitation at 204 nm}.
  {\em Physical Chemistry Chemical Physics}  {\bf 2001}, {\em 3} (15),
  2981--2982.

\bibitem[44]{Fuss2001}
Fu\ss, W.; Schmid, W.; Trushin, S. {Ultrafast electronic relaxation of
  s-trans-butadiene}.  {\em Chemical Physics Letters}  {\bf 2001}, {\em 342}
  (July), 91--98.

\bibitem[45]{Werner1975}
Werner, A.S. {Absolute unimolecular decay rates of energy selected C4H6+
  metastable ions}.  {\em The Journal of Chemical Physics}  {\bf 1975}, {\em
  62} (7) (Apr.), 2900.

\bibitem[46]{Dannacher1980}
Dannacher, J.; Flamme, J.P.; Stadelmann, J.P.; et~al. {Unimolecular
  fragmentations of internal energy selected 1,3-butadiene cations}.  {\em
  Chemical Physics}  {\bf 1980}, {\em 51} (1-2) (Sep.), 189--195.

\bibitem[47]{Fang2011}
Fang, W.; Gong, L.; Zhang, Q.; Shan, X.; Liu, F.; Wang, Z.; et~al.
  {Dissociative photoionization of 1,3-butadiene: experimental and theoretical
  insights.}.  {\em The Journal of Chemical Physics}  {\bf 2011}, {\em 134}
  (17) (May), 174306.

\bibitem[48]{Boguslavskiy2012}
Boguslavskiy, A.E.; Mikosch, J.; Gijsbertsen, A.; Spanner, M.; Patchkovskii,
  S.; Gador, N.; Vrakking, M.J.J.; et~al. The Multielectron Ionization Dynamics
  Underlying Attosecond Strong-Field Spectroscopies.  {\em Science}  {\bf
  2012}, {\em 335} (6074), 1336--1340.

\bibitem[49]{Even2000}
Even, U.; Jortner, J.; Noy, D.; Lavie, N.; et~al. {Cooling of large molecules
  below 1 K and He clusters formation}.  {\em The Journal of Chemical Physics}
  {\bf 2000}, {\em 112} (18), 8068.

\bibitem[50]{Homann2011}
Homann, C.; Krebs, N.; Riedle, E. {Convenient pulse length measurement of
  sub-20-fs pulses down to the deep UV via two-photon absorption in bulk
  material}.  {\em Applied Physics B}  {\bf 2011}, {\em 104} (4) (Aug.),
  783--791.

\bibitem[51]{Parker1997}
Parker, D.H.; Introduction, I. {Velocity map imaging of ions and electrons
  using electrostatic lenses: Application in photoelectron and photofragment
  ion imaging of molecular oxygen}.  {\em Review of Scientific Instruments}
  {\bf 1997}, {\em 68} (September), 3477--3484.

\bibitem[52]{sevi2004}
Osterwalder, A.; Nee, M.J.; Zhou, J.; et~al. High resolution photodetachment
  spectroscopy of negative ions via slow photoelectron imaging.  {\em The
  Journal of Chemical Physics}  {\bf 2004}, {\em 121} (13), 6317--6322.

\bibitem[53]{powis2005}
Garcia, G.A.; Nahon, L.; Harding, C.J.; Mikajlo, E.A.; et~al. A refocusing
  modified velocity map imaging electron/ion spectrometer adapted to
  synchrotron radiation studies.  {\em Review of Scientific Instruments}  {\bf
  2005}, {\em 76} (5), 053302.

\bibitem[54]{hockettThesis2009}
Hockett, P. Photoionization Dynamics of Polyatomic Molecules. Ph.D. thesis,
  University of Nottingham, 2009.

\bibitem[55]{Chichinin2009}
Chichinin, A.I.; Gericke, K.H.; Kauczok, S.; et~al. {Imaging chemical reactions
  – 3D velocity mapping}.  {\em International Reviews in Physical Chemistry}
  {\bf 2009}, {\em 28} (4) (Oct.), 607--680.

\bibitem[56]{Stert1999}
Stert, V.; Radloff, W.; Schulz, C.; et~al. {Ultrafast photoelectron
  spectroscopy: Femtosecond pump-probe coincidence detection of ammonia cluster
  ions and electrons}.  {\em The European Physical Journal D}  {\bf 1999}, {\em
  5} (1), 97.

\bibitem[57]{jochenStats2012}
Mikosch, J. {\em In preperation}  {\bf 2013}.

\bibitem[58]{yang1948}
Yang, C.N. On the Angular Distribution in Nuclear Reactions and Coincidence
  Measurements.  {\em Physical Review}  {\bf 1948}, {\em 74} (7) (Oct),
  764--772.

\bibitem[59]{Reid2003}
Reid, K.L. Photoelectron Angular Distributions.  {\em Annual Review of Physical
  Chemistry}  {\bf 2003}, {\em 54} (1), 397--424.

\bibitem[60]{Dixon1986}
Dixon, R.N. {The determination of the vector correlation between photofragment
  rotational and translational motions from the analysis of Doppler-broadened
  spectral line profiles}.  {\em The Journal of Chemical Physics}  {\bf 1986},
  {\em 85} (4) (Aug.), 1866.

\bibitem[61]{Rakitzis1999}
Rakitzis, T.P.; Zare, R.N. {Photofragment angular momentum distributions in the
  molecular frame: Determination and interpretation}.  {\em The Journal of
  Chemical Physics}  {\bf 1999}, {\em 110} (7) (Feb.), 3341.

\bibitem[62]{Clark2006}
Clark, A.P.; Brouard, M.; Quadrini, F.; et~al. {Atomic polarization in the
  photodissociation of diatomic molecules.}.  {\em Physical Chemistry Chemical
  Physics}  {\bf 2006}, {\em 8} (48) (Dec.), 5591--610.

\bibitem[63]{Suits2008}
Suits, A.G.; Vasyutinskii, O.S. {Imaging atomic orbital polarization in
  photodissociation.}.  {\em Chemical reviews}  {\bf 2008}, {\em 108} (9)
  (Sep.), 3706--46.

\bibitem[64]{Rakitzis2010}
Rakitzis, T.P.; Alexander, A.J. {Photofragment angular momentum distributions
  in the molecular frame. II. Single state dissociation, multiple state
  interference, and nonaxial recoil in photodissociation of polyatomic
  molecules.}.  {\em The Journal of Chemical Physics}  {\bf 2010}, {\em 132}
  (22) (Jun.), 224310.

\bibitem[65]{Chupka1993}
Chupka, W.A. Factors affecting lifetimes and resolution of Rydberg states
  observed in zero-electron-kinetic-energy spectroscopy.  {\em The Journal of
  Chemical Physics}  {\bf 1993}, {\em 98} (6), 4520--4530.

\bibitem[66]{Cockett2005}
Cockett, M.C.R. Photoelectron spectroscopy without photoelectrons: Twenty years
  of ZEKE spectroscopy.  {\em Chemical Society Reviews}  {\bf 2005}, {\em 34}
  (11) (Nov.), 935--948.

\bibitem[67]{White1974}
White, R.; Carlson, T.; Spears, D. {Angular distribution of the photoelectron
  spectra for ethylene, propylene, butene and butadiene}.  {\em Journal of
  Electron Spectroscopy and Related Phenomena}  {\bf 1974}, {\em 3}, 59--70.

\bibitem[68]{Garavelli2001}
Garavelli, M.; Bernardi, F.; Olivucci, M.; Bearpark, M.J.; Klein, S.; et~al.
  {Product Distribution in the Photolysis of s-cis Butadiene: A Dynamics
  Simulation}.  {\em The Journal of Physical Chemistry A}  {\bf 2001}, {\em
  105} (51) (Dec.), 11496--11504.

\bibitem[69]{schuurman2012}
Schuurman, M.  {\em Personal communication}; , 2012.

\bibitem[70]{Fischer1996}
Fischer, I.; Vrakking, M.; Villeneuve, D.; et~al. Femtosecond time-resolved
  zero kinetic energy photoelectron and photoionization spectroscopy studies of
  I-2 wavepacket dynamics.  {\em Chemical Physics}  {\bf 1996}, {\em 207}
  ({2-3}) ({JUL 1}), 331--354.

\bibitem[71]{Wu2011a}
Wu, G.; Boguslavskiy, A.E.; Schalk, O.; Schuurman, M.S.; et~al. {Ultrafast
  non-adiabatic dynamics of methyl substituted ethylenes: the $\pi$3s Rydberg
  state.}.  {\em The Journal of Chemical Physics}  {\bf 2011}, {\em 135} (16)
  (Oct.), 164309.

\bibitem[72]{VanWilderen2011}
van Wilderen, L.J.G.W.; Lincoln, C.N.; van Thor, J.J. {Modelling multi-pulse
  population dynamics from ultrafast spectroscopy.}.  {\em PloS one}  {\bf
  2011}, {\em 6} (3) (Jan.), e17373.

\end{thebibliography}

\end{document}